\documentclass[aps, prd, twocolumn, superscriptaddress, floatfix, longbibliography, nofootinbib]{revtex4-2}

\pdfoutput=1
\usepackage{amsmath}
\usepackage{amssymb}
\usepackage{graphicx}
\usepackage{booktabs}
\usepackage{multirow}
\usepackage{placeins}

\usepackage[colorlinks=true,breaklinks=true]{hyperref}
\usepackage[dvipsnames]{xcolor}
\definecolor{darkred}{rgb}{0.5,0,0}
\definecolor{darkblue}{rgb}{0,0,0.5}
\definecolor{firebrick}{rgb}{0.75,0.125,0.125}
\definecolor{darkgreen}{rgb}{0,0.5,0}
\hypersetup{urlcolor=darkblue,
	    citecolor=darkgreen,
	    linkcolor=firebrick}

\usepackage{orcidlink}

\graphicspath{{figures/}}

\newcommand{\ie}{{\it i.e.}}

\newcommand{\eg}{{\it e.g.}}

\newcommand{\eq}{Eq.}
\newcommand{\eqs}{Eqs.}
\newcommand{\fig}{Fig.}

\providecommand{\Ref}{}
\AtBeginDocument{\renewcommand{\Ref}{Ref.}}
\newcommand{\Refs}{Refs.}
\newcommand{\Tab}{Table}

\newcommand{\equ}[1]{\eq~(\ref{equ:#1})}
\newcommand{\figu}[1]{\fig~\ref{fig:#1}}
\newcommand{\tabl}[1]{\Tab~\ref{tab:#1}}
\newcommand{\Utau}{|U_{\tau 4}|^2}
\newcommand{\Umu}{|U_{\mu 4}|^2}
\newcommand{\Ue}{|U_{e 4}|^2}
\newcommand{\dmsq}{\Delta m^2_{41}}
\newcommand{\fE}[1]{f_{#1,\oplus}}
\newcommand{\fS}[1]{f_{#1,{\rm S}}}
\newcommand{\Pstd}{P^{\rm std}}

\begin{document}

\title{Constraining tau--sterile neutrino mixing with high-energy astrophysical neutrinos}

\author{Mauricio Bustamante\,\orcidlink{0000-0001-6923-0865}}
\email{mbustamante@nbi.ku.dk}
\affiliation{Niels Bohr International Academy, Niels Bohr Institute,\\University of Copenhagen, 2100 Copenhagen, Denmark}

\author{Qinrui Liu\,\orcidlink{0000-0003-3379-6423}}
\email{qinrui\_liu@sfu.ca}
\affiliation{Department of Physics, Simon Fraser University, Burnaby, BC V5A 1S6, Canada}
\affiliation{Arthur B. McDonald Canadian Astroparticle Physics Research Institute, Kingston ON K7L 3N6, Canada}

\date{\today}

\begin{abstract}
Whether a sterile neutrino mixes with the three active ones remains an open question. Its mixing with the tau neutrino, $\Utau$, is the least constrained, since terrestrial experiments probe it only indirectly and over limited ranges of sterile-neutrino mass. We show that TeV--PeV astrophysical neutrinos probe it differently. Their sources make almost no $\nu_\tau$, but oscillations over cosmological distances turn about one third of them into $\nu_\tau$ before they reach Earth. Mixing with a sterile neutrino would make these $\nu_\tau$ scarcer, lowering the $\nu_\tau$ fraction that neutrino telescopes measure, whatever the sterile-neutrino mass. In contrast, sterile mixing with $\nu_e$ or $\nu_\mu$ is nearly undetectable. Fitting the flavor composition measured by IceCube, with the composition at the sources left free, we find $\Utau < 0.88$ at 90\%~C.L. Combining IceCube with present and planned neutrino telescopes, including IceCube-Gen2 and HUNT, could lower this limit to 0.24--0.29 by 2040 and to 0.12--0.14 by 2050. By 2050, these limits would match some present terrestrial limits, though not the strongest.
\end{abstract}

\maketitle

\section{Introduction}
\label{sec:introduction}

Does a fourth neutrino, one that does not feel the weak interaction, mix with the three known ones? Such a sterile neutrino is among the simplest extensions of the Standard Model~\cite{Abazajian:2012ys, Giunti:2019aiy, Diaz:2019fwt, Boser:2019rta, Dasgupta:2021ies, Acero:2022wqg}. Anomalies in short-baseline experiments, LSND~\cite{Aguilar:2001ty}, MiniBooNE~\cite{MiniBooNE:2020pnu}, reactor experiments~\cite{Mention:2011rk}, and gallium experiments~\cite{Kaether:2010ag, SAGE:2009eeu, Giunti:2010zu, Barinov:2021asz}, have pointed to one with a mass splitting $\dmsq \sim 1$~eV$^2$, although global fits find them in tension with other data~\cite{Kopp:2013vaa, Dentler:2018sju}. These anomalies concern the mixing of $\nu_e$ and $\nu_\mu$ with the sterile neutrino, quantified by $|U_{e4}|^2$ and $|U_{\mu4}|^2$, the squared elements of the $4 \times 4$ mixing matrix, $U$, that connect $\nu_e$ and $\nu_\mu$ to the fourth mass eigenstate, $\nu_4$.

\begin{figure}[!t]
  \centering
  \includegraphics[width=\columnwidth]{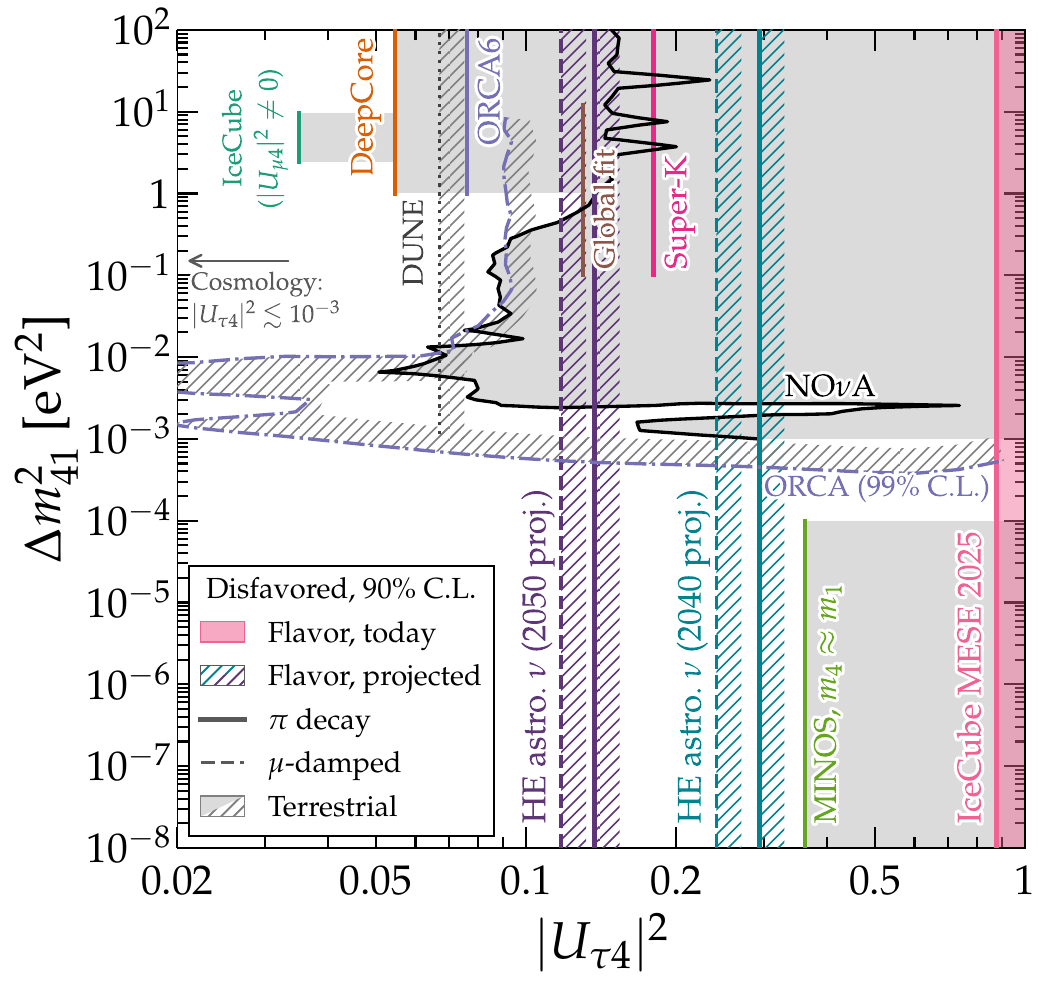}
  \caption{\label{fig:money_plot}\textbf{\textit{Limits on the mixing of $\nu_\tau$ with a sterile neutrino.}}
  Disfavored regions of mixing parameters $\Utau$ and $\dmsq$. Terrestrial limits are from DeepCore~\cite{IceCube:2024dlz}, ORCA6~\cite{KM3NeT:2025iar}, Super-K~\cite{Abe:2014gda}, MINOS~\cite{MINOS:2011ysd}, the global fit of \Ref~\cite{Dentler:2018sju}, NO$\nu$A~\cite{NOvA:2026ksk}, IceCube~\cite{IceCube:2024pky}, and the projections of DUNE~\cite{DUNE:2020fgq} and ORCA~\cite{KM3NeT:2021uez}. The cosmological limit assumes a standard cosmological thermal history~\cite{Hagstotz:2020ukm}. Our limits, from the IceCube flavor composition of high-energy astrophysical neutrinos~\cite{IceCube:2025ole} and from multi-detector projections, hold for any $\dmsq$ and account for the unknown neutrino production mechanism. By 2040, flavor measurements could test $\Utau$ at the level of some terrestrial limits, varying only slightly depending on how neutrinos are made ($\pi$ decay, $\mu$-damped).}
\end{figure}

The mixing of $\nu_\tau$, $\Utau$, is the least tested~\cite{MammenAbraham:2022xoc, Denton:2021mso}. Since no intense $\nu_\tau$ beams exist, terrestrial experiments probe the mixing only indirectly, from its effect on atmospheric $\nu_\mu$ (in Super-K~\cite{Abe:2014gda}, IceCube~\cite{IceCube:2024pky}, IceCube DeepCore~\cite{IceCube:2024dlz}, and KM3NeT/ORCA~\cite{KM3NeT:2025iar}) and accelerator $\nu_\mu$ (in MINOS~\cite{MINOS:2011ysd} and NO$\nu$A~\cite{NOvA:2026ksk}), and only over limited ranges of $\dmsq$. These ranges are set by each experiment's baselines---hundreds to thousands of km---and energies---MeV to TeV. Figure~\ref{fig:money_plot} shows present limits, of $\Utau \lesssim 0.035$ at best. Supernovae constrain $\Utau$ at $\dmsq \gtrsim 10^6$~eV$^2$~\cite{Raffelt:2011nc, Arguelles:2016uwb, Mastrototaro:2019vug, Chauhan:2023sci, Carenza:2023old}, not shown in \figu{money_plot}.

High-energy astrophysical neutrinos offer a test of $\Utau$ that is independent of terrestrial experiments and supernovae, and that holds at every value of $\dmsq$. They are produced at TeV--PeV energies in cosmic accelerators, mainly in the decays of charged pions, and travel cosmological distances before neutrino telescopes detect them~\cite{IceCube:2013low, IceCube:2020wum, IceCube:2025tgp}. Their flavor composition, \ie, how their flux divides among $\nu_e$, $\nu_\mu$, and $\nu_\tau$, depends on how they are made and how they propagate. Hence, measuring it tests both high-energy astrophysics~\cite{Rachen:1998fd, Athar:2000yw, Barenboim:2003jm, Beacom:2003nh, Kashti:2005qa, Mena:2006eq, Kachelriess:2006ksy, Lipari:2007su, Esmaili:2009dz, Choubey:2009jq, Hummer:2010ai, Winter:2013cla, Palladino:2015zua, Bustamante:2015waa, Biehl:2016psj, Bustamante:2019sdb, Ackermann:2019ows, Bustamante:2020bxp, Song:2020nfh, Bhattacharya:2023mmp, Telalovic:2023tcb, Dev:2023znd} and fundamental physics~\cite{Crocker:2001zs, Beacom:2002vi, Barenboim:2003jm, Beacom:2003nh, Beacom:2003eu, Beacom:2003zg, Serpico:2005bs, Mena:2006eq, Lipari:2007su, Pakvasa:2007dc, Esmaili:2009dz, Choubey:2009jq, Esmaili:2009fk, Bhattacharya:2009tx, Bhattacharya:2010xj, Bustamante:2010nq, Mehta:2011qb, Baerwald:2012kc, Fu:2012zr, Pakvasa:2012db, Chatterjee:2013tza, Xu:2014via, Aeikens:2014yga, Arguelles:2015dca, Bustamante:2015waa, Pagliaroli:2015rca, Shoemaker:2015qul, deSalas:2016svi, Gonzalez-Garcia:2016gpq, Bustamante:2016ciw, Rasmussen:2017ert, Dey:2017ede, Bustamante:2018mzu, Farzan:2018pnk, Ahlers:2018yom, Brdar:2018tce, Palladino:2019pid, Ackermann:2019cxh, Arguelles:2019rbn, Ahlers:2020miq, Karmakar:2020yzn, Fiorillo:2020gsb, Song:2020nfh, IceCube:2021tdn, Arguelles:2022tki, MammenAbraham:2022xoc, Telalovic:2023tcb, Telalovic:2025xor}.

What makes these neutrinos sensitive to $\Utau$ is their large $\nu_\tau$ content. On the way to Earth, neutrino oscillations, \ie, changes of flavor caused by mixing and by the differences between the neutrino masses, turn part of the $\nu_e$ and $\nu_\mu$ emitted by the sources into $\nu_\tau$. Under standard mixing, about one third of the neutrinos that reach Earth are expected to be $\nu_\tau$, even though the sources make almost none~\cite{Learned:1994wg}. The non-standard mixing of $\nu_\tau$ with a sterile neutrino would make these $\nu_\tau$ scarcer, lowering their fraction at Earth. The difficulty lies in measuring this, since $\nu_e$ and $\nu_\tau$ make similar signals in the neutrino telescopes that detect them (Sec.~\ref{sec:measurements_flavor}). This makes present measurements of the flavor composition imprecise~\cite{IceCube:2025ole}, but larger detectors, now under construction or planned, could improve them substantially~\cite{Song:2020nfh, Schumacher:2025qca}.

In this work, we infer $\Utau$ from the flavor composition of high-energy astrophysical neutrinos, from its measurements by the IceCube Neutrino Observatory~\cite{IceCube:2025ole} and from projections that combine present and planned neutrino telescopes~\cite{Liu:2023flr, Bustamante:2026aur, Bustamante:2026zst, Barenboim:2026mvn}. Earlier work mapped the flavor composition that sterile mixing can produce at Earth~\cite{Brdar:2016thq, Rasmussen:2017ert, Arguelles:2019tum, Ahlers:2020miq}. We work in reverse, fitting active-sterile mixing to flavor measurements in order to constrain it. Of the three active-sterile mixings, only $\Utau$ has an effect that flavor measurements can detect in spite of the unknown flavor composition at the sources.

Figure~\ref{fig:money_plot} summarizes our limits on $\Utau$, contrasted with terrestrial experiments. \textbf{Today, IceCube disfavors only large mixing, $\Utau > 0.88$ at 90\%~confidence level (C.L.). Future multi-detector  measurements could tighten this limit to the level of some of today's terrestrial limits.} Over the ranges of $\dmsq$ in which terrestrial limits are quoted, our projected limits for 2050 would be stronger than those of Super-K and MINOS, but weaker than those from atmospheric neutrinos at the eV scale and from NO$\nu$A, and than the limit expected from DUNE. Still, they complement terrestrial limits in two ways. First, they hold at small values of $\dmsq$, where most terrestrial experiments quote no limit. Second, they rest on systematic uncertainties different from those of terrestrial experiments.

Section~\ref{sec:flavor} shows how a fourth neutrino changes the flavor composition at Earth and why only $\Utau$ leaves a detectable mark. Sections~\ref{sec:measurements} and \ref{sec:statistics} describe the flavor measurements that we use and our statistical method. Section~\ref{sec:results} presents our results, which Sec.~\ref{sec:summary} summarizes. The appendices contain the projected likelihoods (Appendix~\ref{app:likelihoods}), further statistical checks (Appendix~\ref{app:statistics}), the derivations behind Sec.~\ref{sec:flavor} (Appendices~\ref{app:single_angle}--\ref{app:emu}), and the profiles of the three active-sterile mixings (Appendix~\ref{app:mixings}).

\section{Neutrino flavor composition}
\label{sec:flavor}

\subsection{Standard and 3+1 mixing}
\label{sec:flavor_mixing}

With standard mixing, the three active flavors, $\nu_e$, $\nu_\mu$, and $\nu_\tau$, are superpositions of three mass eigenstates, $\nu_1$, $\nu_2$, and $\nu_3$, with masses $m_1$, $m_2$, and $m_3$. This mixing is set by the angles $\theta_{12}$, $\theta_{23}$, and $\theta_{13}$ and by a CP-violation phase, $\delta_{\rm CP}$. Because the mass eigenstates acquire different phases as they travel, a neutrino produced as one flavor can later be detected as another. The lengths over which these flavor oscillations develop are set by the mass splittings, $\Delta m^2_{ij} \equiv m_i^2 - m_j^2$, of which two are independent, $\Delta m^2_{21}$ and $\Delta m^2_{31}$. Global fits to oscillation data constrain these parameters. We use their values from NuFIT~6.1~\cite{Esteban:2024eli, NuFIT61}. For our main results, we adopt normal mass ordering, $\Delta m^2_{31} > 0$; we test inverted ordering, $\Delta m^2_{31} < 0$, separately (\tabl{parameters}).

We consider the 3+1 scenario, in which a sterile neutrino, $\nu_s$, which has no weak interactions, joins the three active flavors; for reviews, see \Refs~\cite{Abazajian:2012ys, Giunti:2019aiy, Diaz:2019fwt, Boser:2019rta, Dasgupta:2021ies, Acero:2022wqg}. The scenario adds a fourth mass eigenstate, $\nu_4$, with mass $m_4$. Of the three mass splittings that involve $\nu_4$, only one is independent (the other two follow from it and the standard splittings). We take it to be $\dmsq$, so that $\Delta m^2_{42} = \dmsq - \Delta m^2_{21}$ and $\Delta m^2_{43} = \dmsq - \Delta m^2_{31}$.

A $4 \times 4$ unitary matrix, $U$, connects the four flavor states to the four mass eigenstates. We parametrize it, with the ordering of rotations of \Ref~\cite{Dentler:2018sju}, as
\begin{equation}
  \label{equ:mixing}
  U = R_{34} R_{24} R_{14} R_{23} R_{13} R_{12} ,
\end{equation}
where $R_{ij}$ is a rotation by the angle $\theta_{ij}$ in the $i$-$j$ plane; the rotations $R_{13}$, $R_{14}$, and $R_{24}$ also carry the CP-violation phases $\delta_{\rm CP}$, $\delta_{14}$, and $\delta_{24}$. When the three new angles, $\theta_{14}$, $\theta_{24}$, and $\theta_{34}$, are zero, $U$ reduces to the standard Pontecorvo-Maki-Nakagawa-Sakata mixing matrix. The active-sterile mixings, \ie, the active-flavor content of $\nu_4$, are $\Ue = s_{14}^2$, $\Umu = c_{14}^2 s_{24}^2$, and $\Utau = c_{14}^2 c_{24}^2 s_{34}^2$, where $s_{ij} \equiv \sin\theta_{ij}$ and $c_{ij} \equiv \cos\theta_{ij}$. These mixings are the parameters that we constrain.

\subsection{High-energy astrophysical neutrinos}
\label{sec:flavor_averaged}

High-energy astrophysical neutrinos, with TeV--PeV energies, are thought to be made in cosmic accelerators, where high-energy protons collide with matter or radiation and produce charged pions. The decay chain $\pi^+ \to \mu^+ + \nu_\mu$, $\mu^+ \to e^+ + \nu_e + \bar\nu_\mu$, together with its charge conjugate, yields one $\nu_e$ for every two $\nu_\mu$, counting neutrinos and antineutrinos together (since high-energy neutrino telescopes do not ordinarily tell them apart). This ratio corresponds to a flavor composition at the source, \ie, a fraction of each flavor, of $(f_{e,{\rm S}}, f_{\mu,{\rm S}}, f_{\tau,{\rm S}}) = (1/3, 2/3, 0)$~\cite{Learned:1994wg}. When the muons cool via synchrotron emission before they decay, only the $\nu_\mu$ from the pion decay keep high energies, so the composition becomes $(0, 1, 0)_{\rm S}$; we call such sources muon-damped~\cite{Kashti:2005qa}. Neutron decay yields $(1, 0, 0)_{\rm S}$ but is unlikely to dominate, since its $\bar\nu_e$ take a much smaller share of the parent energy than in pion decay, so we do not consider it further.

All of these compositions have the form $(\fS{e}, 1 - \fS{e}, 0)$, which we adopt for our analysis (Sec.~\ref{sec:results_limits} allows for up to 10\% $\nu_\tau$ content, too). Since the sources and production mechanism of high-energy neutrinos are unknown, we leave $\fS{e}$ free in our statistical analysis. Projecting a future measurement, however, requires assuming a flavor composition at the sources. Our projections assume pion-decay or muon-damped sources, although our fits to them still leave $\fS{e}$ free (Sec.~\ref{sec:statistics_likelihood}).

Astrophysical neutrinos travel distances, $L$, of Mpc to Gpc, far longer than their oscillation lengths, $L^{\rm osc}_{ij} = 4\pi E / |\Delta m^2_{ij}|$, where $E$ is the neutrino energy. At 100~TeV, the longest of the standard oscillation lengths, set by $\Delta m^2_{21}$, is about $10^{-4}$~pc. Because the detected neutrinos span a range of energies and distances, the oscillations average out~\cite{Learned:1994wg}. For $\nu_4$, averaging requires $L \gg L^{\rm osc}_{41}$, \ie, $\dmsq \gg 4\pi E/L \approx 8 \times 10^{-17}~{\rm eV}^2 \, (E/{\rm PeV}) ({\rm Gpc}/L)$, which every value of $\dmsq$ in \figu{money_plot} satisfies by many orders of magnitude. This, combined with the limited energy resolution of neutrino telescopes, makes them sensitive to the average flavor-transition probability.  For a neutrino produced as $\nu_\alpha$ to be detected as a $\nu_\beta$, this is
\begin{equation}
  \label{equ:prob}
  P_{\alpha\beta} = \sum_{i=1}^{4} |U_{\alpha i}|^2 |U_{\beta i}|^2 .
\end{equation}

The fraction of neutrinos that reach Earth as $\nu_\beta$ is
$F_\beta = \sum_{\alpha=e,\mu,\tau,s} \fS{\alpha} P_{\alpha\beta}$. Because every neutrino reaches Earth as one of the four flavors, $F_e + F_\mu + F_\tau + F_s = 1$. However, neutrino telescopes detect only active flavors, so the flavor fractions that they measure are $F_e$, $F_\mu$, and $F_\tau$, rescaled to sum to one (see, \eg, \Ref~\cite{Ahlers:2020miq}):
\begin{equation}
  \label{equ:fearth}
  \fE{\beta} = \frac{F_\beta}{\sum_{\gamma=e,\mu,\tau} F_\gamma} .
\end{equation}

Since the fractions in \equ{fearth} are normalized over the active flavors, they sum to one even when sterile mixing turns part of the flux into $\nu_s$. With standard mixing, at the best-fit values of NuFIT~6.1~\cite{Esteban:2024eli, NuFIT61}, pion decay yields $(\fE{e}, \fE{\mu}, \fE{\tau}) \approx (0.33, 0.34, 0.33)$; muon-damped sources, $(0.22, 0.41, 0.38)$; and neutron decay, $(0.55, 0.22, 0.24)$. We show each composition as a point in the flavor triangle, a ternary plot whose three coordinates are the flavor fractions and whose vertices are pure $\nu_e$, $\nu_\mu$, and $\nu_\tau$; see Figs.~\ref{fig:mixing_directions} and \ref{fig:regions}. As $\fS{e}$ varies from 0 to 1, the composition at Earth moves along a line in the triangle, the segment of standard compositions, which joins the muon-damped and the neutron-decay compositions.  The uncertainties of the standard mixing parameters widen this segment into a band, the standard-mixing region, which contains the compositions that sources $(\fS{e}, 1 - \fS{e}, 0)$ can reach with standard mixing~\cite{Bustamante:2015waa, Song:2020nfh, Liu:2023flr}, shown at 90\%~C.L.~in Figs.~\ref{fig:mixing_directions} and \ref{fig:regions}.

\subsection{Sensitivity to each active-sterile mixing}
\label{sec:flavor_see}

\begin{figure}[t]
  \centering
  \includegraphics[width=\columnwidth]{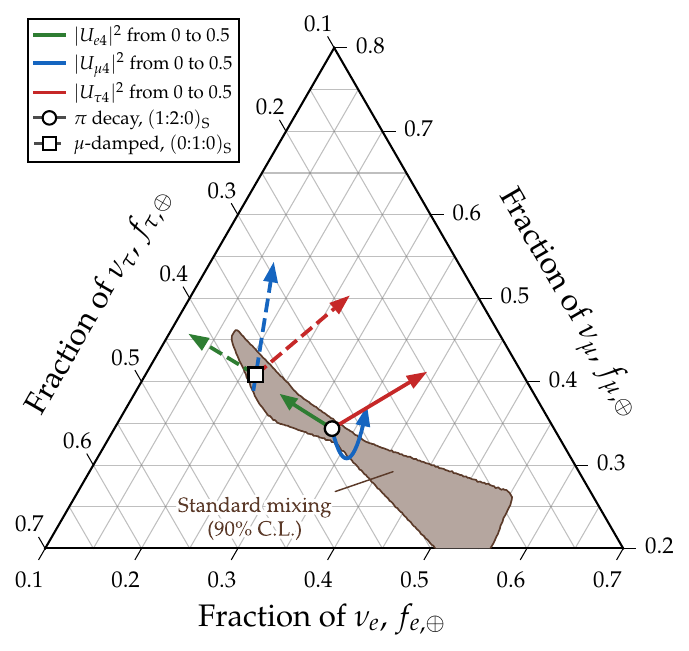}
  \caption{\label{fig:mixing_directions}\textbf{\textit{How sterile mixing shifts flavor composition.}} Flavor composition at Earth as one active-sterile mixing grows, with the other two at zero, starting from compositions yielded by standard mixing (NuFIT~6.1 best fit~\cite{Esteban:2024eli, NuFIT61}). The band is the standard-mixing region of \figu{regions}. Sterile mixing with $\nu_e$ moves the composition along the band; sterile mixing with $\nu_\mu$ moves it away and back; only sterile mixing with $\nu_\tau$ moves it steadily away. See Sec.~\ref{sec:flavor_see} for details.}
\end{figure}

We examine each active-sterile mixing alone, to show why flavor measurements are most sensitive to $\Utau$, the only nonzero active-sterile mixing in our default analysis (Sec.~\ref{sec:statistics_likelihood}). When only one active-sterile mixing angle, $\theta_{x4}$ with $x = e$, $\mu$, or $\tau$, is nonzero, the probabilities of \equ{prob} take a simple form (Appendix~\ref{app:single_angle}),
\begin{equation}
  \label{equ:single_angle}
  P_{\alpha\beta} = c_\alpha^2 c_\beta^2 \, \Pstd_{\alpha\beta} + \delta_{\alpha x} \delta_{\beta x} \, s_x^4 ,
\end{equation}
where $\Pstd_{\alpha\beta}$ is the standard-mixing probability, $c_\alpha^2 \equiv \cos^2\theta_{x4} = (1-|U_{x4}|^2)$ for $\alpha = x$ and $c_\alpha \equiv 1$ otherwise, $s_x \equiv \sin\theta_{x4} = |U_{x4}|$, and $\delta_{\alpha x}$ is the Kronecker delta. The factor $c_\alpha^2$ lowers all $P_{x\beta}$, the probabilities for a neutrino emitted as $\nu_x$, and $c_\beta^2$ lowers all $P_{\alpha x}$, those for a neutrino detected as $\nu_x$. The term $s_x^4$ raises the probability that a neutrino emitted as a $\nu_x$ is detected as a $\nu_x$.

Since $c_\alpha^2$ lowers all $P_{x\beta}$ by the same factor, it has the same effect as a source that emits fewer $\nu_x$, which our analysis accounts for by leaving $\fS{e}$ free. Our analysis uses the exact probabilities of \equ{prob}, but, to illustrate how active-sterile mixing shifts the flavor composition (Appendix~\ref{app:directions}), we take the mixing to be small, so the term $s_x^4$ in \equ{single_angle} is negligible. The flavor shift is then
\begin{equation}
  \label{equ:directions}
  \delta \fE{\beta} = - |U_{x4}|^2 \, \fE{x} \left( \delta_{\beta x} - \fE{\beta} \right) + \mathcal{O}(|U_{x4}|^4) ,
\end{equation}
proportional to $\boldsymbol{f}_\oplus - \boldsymbol{e}_x$, where $\boldsymbol{f}_\oplus = (\fE{e}, \fE{\mu}, \fE{\tau})$ and $\boldsymbol{e}_x$ is the vertex of the flavor triangle that corresponds to flavor $x$ [\eg, $\boldsymbol{e}_e = (1, 0, 0)$]. The composition moves along the line that joins it to that vertex, away from the vertex, by a fraction $|U_{x4}|^2 \fE{x}$ of the distance between them. For $x = \tau$, the composition moves along this line for any value of $\Utau$ (Sec.~\ref{sec:flavor_tau}).

Figure~\ref{fig:mixing_directions} shows how the composition moves as each active-sterile mixing grows from zero, starting from the compositions that pion-decay and muon-damped sources yield with standard mixing. Since varying $\fS{e}$ moves the composition along the long axis of the standard-mixing region (Sec.~\ref{sec:flavor_averaged}), a change of source---\ie, of $\fS{e}$---can absorb the part of a shift along this axis, but not the part perpendicular to it. At the pion-decay composition, raising $\Ue$, $\Umu$, or $\Utau$ from zero moves the composition in a direction $3.4^\circ$, $42^\circ$, or $66^\circ$ from the long axis, respectively. For equal small active-sterile mixings, the part of the shift perpendicular to the long axis is more than ten times smaller for $\nu_e$ than for $\nu_\mu$ or $\nu_\tau$.

\subsubsection{Sterile mixing with $\nu_e$}

Sterile mixing with $\nu_e$ lowers all $P_{e\beta}$ and all $P_{\alpha e}$. Lowering the $P_{e\beta}$ has the same effect as a source with a different $\fS{e}$, which our analysis absorbs because $\fS{e}$ is free (Appendix~\ref{app:emu}). Lowering the $P_{\alpha e}$ moves the composition away from the $\nu_e$ vertex [\equ{directions}]. At the pion-decay composition, this direction is closely aligned to the long axis of the standard-mixing region (Appendix~\ref{app:directions}). Hence, a change in $\fS{e}$ can also absorb most of this effect. As a result, flavor measurements leave $\Ue$ almost unconstrained (\figu{profiles}). At its terrestrial limit, $\Ue = 0.017$~\cite{Blennow:2025qgd}, sterile mixing with $\nu_e$ would shift each flavor fraction by at most 0.005 even for a fixed source, far below the projected measurement precision.

\subsubsection{Sterile mixing with $\nu_\mu$}

Sterile mixing with $\nu_\mu$ also lowers all $P_{\mu\beta}$ and all $P_{\alpha\mu}$. Unlike for $\nu_e$, lowering the $P_{\alpha\mu}$ moves the composition at a large angle to the long axis of the standard-mixing region, so a change in $\fS{e}$ cannot mimic it. As $\Umu$ grows, however, the additive term $|U_{\mu4}|^4$ in \equ{single_angle} offsets this shift (\figu{mixing_directions}). The compensation is complete at $\Umu_\star = f^{\rm std}_{\mu,\oplus} / (\fS{\mu} + f^{\rm std}_{\mu,\oplus})$, which is 0.34 for pion decay and 0.29 for muon-damped sources (Appendix~\ref{app:umu}). At $\Umu_\star$, a source with a suitable $\fS{e}$ yields exactly the standard composition, so flavor measurements can neither discover nor disfavor values of $\Umu$ near it (Sec.~\ref{sec:results_other} and Appendix~\ref{app:statistics_reach}). At its terrestrial limit, $\Umu = 0.024$~\cite{Blennow:2025qgd}, sterile mixing with $\nu_\mu$ would shift each flavor fraction by at most 0.006, far below the projected measurement precision.

\subsubsection{Sterile mixing with $\nu_\tau$}
\label{sec:flavor_tau}

\begin{figure}[t]
  \centering
  \includegraphics[width=\columnwidth]{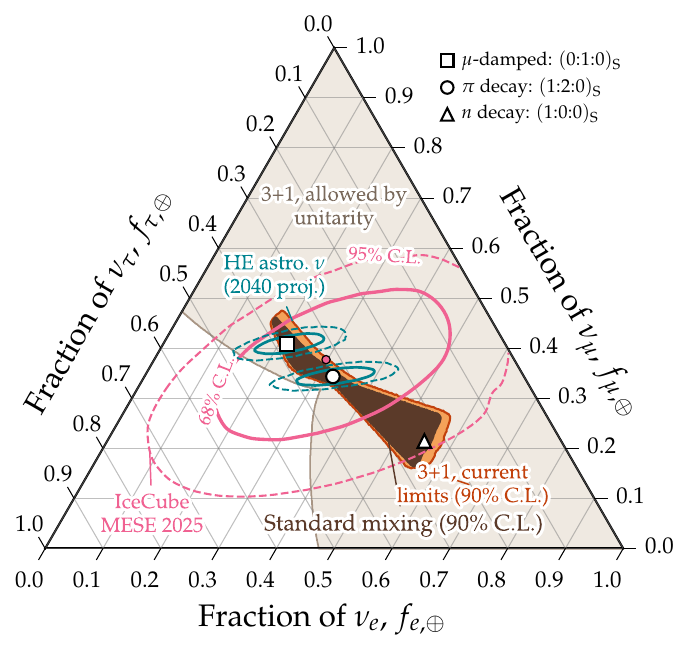}
  \caption{\label{fig:regions}\textbf{\textit{Flavor composition at Earth with standard and with 3+1 mixing.}}
  Regions at 90\%~C.L., for sources $(\fS{e}, 1 - \fS{e}, 0)$ with any $\fS{e} \in [0, 1]$: with standard mixing, from NuFIT~6.1~\cite{Esteban:2024eli, NuFIT61}; with 3+1 mixing within today's eV-scale limits, $\Ue < 0.017$, $\Umu < 0.024$~\cite{Blennow:2025qgd}, and $\Utau < 0.057$, with $\Umu$ free~\cite{IceCube:2024dlz}; and with any unitary 3+1 mixing~\cite{Ahlers:2020miq}. The measurements are IceCube MESE~2025~\cite{IceCube:2025ole} and the main 2040 projection (\tabl{detectors}). Sterile mixing within today's limits barely enlarges the standard region. See Sec.~\ref{sec:flavor_tau} for details.}
\end{figure}

Sterile mixing with $\nu_\tau$ also lowers all $P_{\tau\beta}$ and all $P_{\alpha\tau}$. Since the sources emit no $\nu_\tau$, only lowering the $P_{\alpha\tau}$ matters: for $\alpha = e, \mu$, $P_{\alpha\tau} = (1 - \Utau) \Pstd_{\alpha\tau}$, while $P_{\alpha e}$ and $P_{\alpha\mu}$ keep their standard values [\equ{single_angle}]. Hence, the ratio $\fE{e} : \fE{\mu}$ keeps its standard value, while $\fE{\tau}$ falls (Appendix~\ref{app:tau_deficit}): sterile mixing with $\nu_\tau$ creates a $\nu_\tau$ deficit. To first order, $\fE{\tau}$ falls by $\Utau \fE{\tau} (1 - \fE{\tau})$ [\equ{directions}]; \eg, for pion decay, $\Utau = 0.3$ lowers it from 0.33 to 0.26. With $\fE{e} : \fE{\mu}$ fixed, the composition moves along a straight line away from the $\nu_\tau$ vertex, at a large angle to the long axis of the standard-mixing region (\figu{mixing_directions}). No change of $\fS{e}$ can absorb this shift, which is why we focus on $\Utau$. Sources where charmed mesons decay also emit $\nu_\tau$, from the decays of $D_s^\pm$ mesons; for neutrinos from charm decays alone, $\fS{\tau} \approx 0.05$~\cite{Enberg:2008te}. These $\nu_\tau$ partly offset the deficit, but do not invalidate our projected limits on $\Utau$, which we compute for $\fS{\tau}$ up to 0.10 (Sec.~\ref{sec:results_limits}).

Figure~\ref{fig:regions} shows the standard-mixing region and two regions that the same sources reach with 3+1 mixing. For the first, we vary all three active-sterile mixings within today's eV-scale limits, along with the standard mixing parameters. The second, from \Ref~\cite{Ahlers:2020miq}, allows any unitary $4 \times 4$ mixing matrix. Sterile mixing within today's eV-scale limits widens the standard-mixing region only slightly, mostly toward lower $\fE{\tau}$. The IceCube MESE~2025 flavor measurement~\cite{IceCube:2025ole} cannot yet single out sterile mixing, since its 95\%~C.L.~contour contains nearly all of the standard-mixing region. The uncertainty of our main 2040 projection on $\fE{\tau}$ (\tabl{detectors}) is comparable to the $\nu_\tau$ deficit that $\Utau \approx 0.3$ produces.

Since the sources emit no $\nu_4$ ($U_{e4} = U_{\mu4} = 0$), the $\nu_\tau$ deficit at Earth does not depend on the mass of $\nu_4$, even if $\nu_4$ is too heavy to be produced (Appendix~\ref{app:tau_deficit}). Terrestrial limits, in contrast, hold only over specific ranges of $\dmsq$ (Sec.~\ref{sec:results_terrestrial}). Inside the Earth, the deficit does depend on $\dmsq$. Matter there acts on the active neutrinos but not on $\nu_s$, so it drives $\nu_\tau$--$\nu_s$ conversions when $\dmsq/(2E)$ is comparable to the matter potential~\cite{Nunokawa:2003ep, IceCube:2016rnb, IceCube:2024pky}. Our analysis leaves out this effect, because the flavor measurements that we use assume standard propagation inside the Earth. We estimate its size separately, for an isotropic flux $\propto E^{-2.5}$ from 20~TeV to 10~PeV, averaged over directions and energies. For $\Utau = 0.1$--0.3, it changes the $\nu_\tau$ deficit by up to about 15\% for $\dmsq \approx 3$--30~eV$^2$, and by less than 5\% outside 0.5--100~eV$^2$. It can deepen or reduce the deficit, depending on the ratio of neutrinos to antineutrinos. The limits of Sec.~\ref{sec:results_limits} do not include this Earth-matter effect.

\section{Flavor measurements}
\label{sec:measurements}

\subsection{Detecting high-energy neutrinos}
\label{sec:measurements_detection}

Neutrino telescopes detect the Cherenkov light of the charged particles that high-energy neutrinos make when they interact with nucleons, in ice, as in IceCube~\cite{IceCube:2016zyt}, or in water, as in KM3NeT~\cite{KM3Net:2016zxf} and Baikal-GVD~\cite{Baikal-GVD:2025rhg}. A charged-current (CC) interaction produces a charged lepton of the same flavor as the neutrino; a neutral-current (NC) interaction produces a neutrino. In both, the nucleon breaks up into a shower of hadrons. The amount, timing, and pattern of the light reveal the energy and direction of the neutrino and the topology of the event.

Each flavor makes a different mix of event topologies. In CC interactions, a $\nu_\mu$ makes a muon that leaves a kilometers-long \textit{track} of light, while a $\nu_e$ makes an electron that deposits its energy within a few meters, in a nearly spherical \textit{shower}; NC interactions of all flavors make showers, too. A $\nu_\tau$ makes a tau that travels about 50~m per PeV before decaying, usually into a second shower; above a few hundred TeV, the two showers can be resolved as a \textit{double cascade}~\cite{Learned:1994wg, IceCube:2020fpi}. In 17\% of its decays, the tau makes a muon, which leaves a track. A $\bar\nu_e$ of 6.3~PeV can also produce a $W$ boson resonantly on an electron, the Glashow resonance~\cite{Glashow:1960zz, IceCube:2021rpz}.

\subsection{From events to flavor composition}
\label{sec:measurements_flavor}

The numbers of tracks, showers, and double cascades in a sample of events measure the flavor composition at Earth~\cite{Beacom:2003nh, Bustamante:2015waa}. The share of tracks measures mainly $\fE{\mu}$. In contrast, $\nu_e$ and $\nu_\tau$ both make mostly similar showers. Only double cascades and Glashow-resonance events, which are rare, tell them apart: IceCube has found two double cascades in 7.5~years of high-energy starting events~\cite{IceCube:2020fpi}, seven $\nu_\tau$ candidates in 9.7~years with a dedicated  search~\cite{IceCube:2024nhk}, and one Glashow-resonance event in 4.6~years~\cite{IceCube:2021rpz}. Therefore, flavor measurements carry a $\nu_e$--$\nu_\tau$ degeneracy: they constrain $\fE{\mu}$ more tightly than the split between $\fE{e}$ and $\fE{\tau}$.

This degeneracy limits the sensitivity to $\Utau$. Sterile mixing with $\nu_\tau$ lowers $\fE{\tau}$ and raises $\fE{e}$ and $\fE{\mu}$ [\equ{directions}]. A source with more $\nu_e$ can restore $\fE{\mu}$, but then leaves $\fE{e}$ higher and $\fE{\tau}$ lower than their standard values. Thus, the change that remains is in the split between $\fE{e}$ and $\fE{\tau}$, the least precise part of a flavor measurement. A technique that separates $\nu_e$ from $\nu_\tau$ better would tighten the limits on $\Utau$ (Sec.~\ref{sec:measurements_projections}).

A flavor measurement compares the detected events, sorted by topology and binned in reconstructed energy and direction, with those expected from the astrophysical flux, which depend on $\boldsymbol{f}_\oplus$, and from the backgrounds of atmospheric neutrinos and muons~\cite{IceCube:2015rro, IceCube:2015gsk, IceCube:2020wum, IceCube:2025ole}. Profiling the nuisance parameters, such as the normalization and spectral shape of the astrophysical flux, yields the test statistic $\mathrm{TS}(\boldsymbol{f}_\oplus) = -2 \ln [\mathcal{L}(\boldsymbol{f}_\oplus) / \mathcal{L}_{\max}]$, where $\mathcal{L}$ is the profile likelihood and $\mathcal{L}_{\max}$ its maximum. Since flavor fractions sum to one, the TS depends on two of them; we take $\fE{e}$ and $\fE{\tau}$. The TS is all that our analysis takes from each measurement (Sec.~\ref{sec:statistics}); \Refs~\cite{Bustamante:2019sdb, Song:2020nfh} describe its construction for present and future telescopes.

\subsection{Present flavor measurement}
\label{sec:measurements_present}

The most recent flavor measurement, IceCube MESE~2025~\cite{IceCube:2025ole}, uses 9\,888 medium-energy starting events (MESE), \ie, events whose interaction vertex lies inside the detector, collected over 11.4~years, with energies from 5~TeV to 10~PeV. Its best-fit composition, $(\fE{e}, \fE{\mu}, \fE{\tau}) = (0.30, 0.37, 0.33)$, agrees with that from pion-decay sources with standard mixing, $(0.33, 0.34, 0.33)$. It is the first measurement to find all three flavor fractions nonzero at more than 90\%~C.L. As the $\nu_e$--$\nu_\tau$ degeneracy implies, $\fE{\tau} = 0.33^{+0.24}_{-0.25}$ is less constrained than $\fE{\mu} = 0.37^{+0.14}_{-0.16}$.

IceCube has made public the TS of this measurement on a grid of $(\fE{e}, \fE{\tau})$~\cite{IceCube:2026mesedata}, for two models of the energy spectrum of the astrophysical flux: a single power law and a broken power law. We use the broken power law, which the data favor over the single one at more than $4\sigma$~\cite{IceCube:2025tgp}. (With the single power law, our limit on $\Utau$ derived in Sec.~\ref{sec:results_current} would tighten only slightly.)

\subsection{Projected flavor measurements}
\label{sec:measurements_projections}

For the projected measurements, we adopt the forecasts of \Refs~\cite{Liu:2023flr, Bustamante:2026aur, Bustamante:2026zst, Barenboim:2026mvn}, translated onto the flavor compositions expected with NuFIT~6.1, and build a flavor-composition likelihood from the posterior samples of each (Appendix~\ref{app:likelihoods}). Like \Ref~\cite{IceCube:2015gsk}, these forecasts combine two event samples. High-Energy Starting Events (HESE) have their interaction vertex inside the detector~\cite{Schonert:2008is, IceCube:2013low, Gaisser:2014bja, IceCube:2014stg, Arguelles:2018awr, IceCube:2020wum}. Through-going muons, made by $\nu_\mu$ that interact outside the detector, reveal only the $\nu_\mu$ content of the flux, but are far more numerous~\cite{IceCube:2019cia, IceCube:2021xar}. The forecasts combine the telescopes that operate today, IceCube, Baikal-GVD, and KM3NeT, with planned ones~\cite{MammenAbraham:2022xoc, Ackermann:2022rqc}: P-ONE~\cite{P-ONE:2020ljt}, IceCube-Gen2~\cite{IceCube-Gen2:2020qha}, NEON~\cite{Zhang:2024slv}, TRIDENT~\cite{TRIDENT:2022hql}, and HUNT~\cite{Huang:2023mzt}, the largest of them about 30 times the size of IceCube. Each telescope is treated as a copy of IceCube whose event rate scales with its size~\cite{Schumacher:2025qca} (see also \Refs~\cite{Song:2020nfh, Fiorillo:2022rft, Telalovic:2023tcb, Liu:2023flr}).

\begin{figure}[t]
  \centering
  \includegraphics[width=\columnwidth]{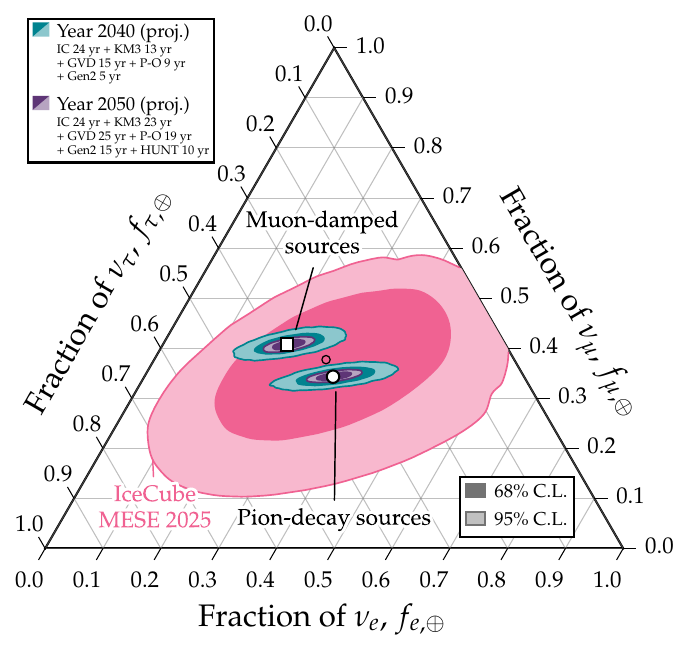}
  \caption{\label{fig:likelihoods}\textbf{\textit{Present and projected flavor-composition likelihoods.}}
  IceCube MESE~2025~\cite{IceCube:2025ole} and the main multi-detector projections (\tabl{detectors}), each projection centered on the flavor composition that its source yields with standard mixing (Appendix~\ref{app:likelihoods}). The regions are elongated along lines of constant $\fE{\mu}$, because showers from $\nu_e$ and $\nu_\tau$ look alike (Sec.~\ref{sec:measurements_flavor}). The projected measurements are far more precise than the present one, most of all in $\fE{\mu}$.}
\end{figure}

We consider eight projected detector combinations, listed in \tabl{limits} and, with the years of data from each detector, in \tabl{detectors}. The detectors start taking data between 2011, for IceCube, and 2040, for TRIDENT, NEON, and HUNT. In the combinations with IceCube-Gen2, IceCube stops in 2035, when IceCube-Gen2 replaces it. By 2050, the combined measurement will resolve departures of order 5\% of each flavor fraction from its standard value. Our main projections come from IceCube, KM3NeT, Baikal-GVD, P-ONE, and IceCube-Gen2 by 2040, and from these five plus HUNT by 2050.

Figure~\ref{fig:likelihoods} shows the flavor-composition likelihoods of IceCube MESE~2025 and of the main projections, assuming neutrino production via full and muon-damped pion decay; Appendix~\ref{app:likelihoods} shows the other projections. Because of the $\nu_e$--$\nu_\tau$ degeneracy, the allowed regions are elongated roughly along lines of constant $\fE{\mu}$; at 68\%~C.L., those of the main projections are four to five times longer than wide. A projection can detect the $\nu_\tau$ deficit once the deficit exceeds the width of its allowed region (Sec.~\ref{sec:results}).

Better identification of $\nu_\tau$ would measure $\fE{\tau}$ more precisely. One technique is already in use: IceCube MESE~2025 identifies double cascades above 30~TeV~\cite{IceCube:2025ole}, half the energy threshold of the earlier analysis of high-energy starting events~\cite{IceCube:2020fpi}. Two others are under study. The first separates the electromagnetic showers of $\nu_e$ from the mostly hadronic showers of $\nu_\tau$ through the delayed Cherenkov light, or ``echoes'', emitted after the shower by decaying muons and captured neutrons~\cite{Li:2016kra}; studies of the neutron-capture echo in IceCube show promising preliminary results~\cite{Steuer:2017tca, Farrag:2023jut, IceCube:2025paj}. The second uses starting tracks, since those from tau decays deposit more of their energy at the interaction vertex than those from $\nu_\mu$; with existing IceCube data, it already gives a competitive sensitivity to the $\nu_\tau$-to-$\nu_\mu$ flux ratio~\cite{Wen:2026ngx}. Our projections include neither technique, which makes them, and the limits that we derive from them, conservative.

\section{Statistical methods}
\label{sec:statistics}

We adopt a frequentist approach: a profile likelihood compares the flavor composition at Earth predicted with 3+1 mixing to each flavor measurement, with the standard mixing parameters and the source composition as nuisance parameters. From it, we derive upper limits on $\Utau$ and, for the projections, the significance with which standard mixing could be rejected, and we check their confidence levels with pseudo-experiments. The projected measurements come from Bayesian forecasts (Sec.~\ref{sec:measurements_projections}). Because these use uniform priors on the flavor composition, their posteriors are proportional to the likelihoods that our analysis needs (Appendix~\ref{app:likelihoods}).

\subsection{The likelihood}
\label{sec:statistics_likelihood}

The flavor composition at Earth depends on three sets of parameters: the $\nu_e$ fraction at the sources, $\fS{e}$; the active-sterile mixings, $\boldsymbol{U} \equiv (\Ue, \Umu, \Utau)$; and the standard mixing parameters, $\boldsymbol{\vartheta} \equiv (s_{12}^2, s_{23}^2, s_{13}^2, \delta_{\rm CP})$. For sources $(\fS{e}, 1 - \fS{e}, 0)$, \eqs~(\ref{equ:prob}) and (\ref{equ:fearth}) give it as $\boldsymbol{f}_\oplus(\fS{e}; \boldsymbol{U}, \boldsymbol{\vartheta})$. We compare it with a flavor measurement through the TS of Sec.~\ref{sec:measurements_flavor} and add pull terms for the parameters that other experiments constrain:
\begin{equation}
  \label{equ:chi2}
  \begin{aligned}
    \chi^2(\boldsymbol{U}, \boldsymbol{\vartheta}) = {} & \min_{0 \le \fS{e} \le 1} \mathrm{TS}\!\left[ \boldsymbol{f}_\oplus(\fS{e}; \boldsymbol{U}, \boldsymbol{\vartheta}) \right] \\
    & + \chi^2_{\rm std}(\boldsymbol{\vartheta}) + \chi^2_{\rm terr}(\boldsymbol{U}) .
  \end{aligned}
\end{equation}

\tabl{parameters} lists the parameters of the fits. The parameter of interest is $\Utau$; all others are nuisance parameters, which we profile, \ie, over which we minimize $\chi^2$, either freely or subject to the pull terms. The default analysis makes three assumptions: the only nonzero active-sterile mixing is $\Utau$; the sources emit neither $\nu_\tau$ nor $\nu_s$; and the mass ordering is normal. The variants of Sec.~\ref{sec:results_limits} relax them one at a time; when $\Ue$ or $\Umu$ is free, we also profile over the phase $\delta_{14}$ or $\delta_{24}$. The mass splitting $\dmsq$ is not a parameter of the fits, because the averaged probabilities, \equ{prob}, do not depend on it (Sec.~\ref{sec:flavor_tau}).

\begin{table}[t]
  \caption{\label{tab:parameters}\textbf{\textit{Parameters of the fits.}} The parameter of interest and the nuisance parameters, in the default analysis and in the variants that test its robustness, each of which changes one assumption (Sec.~\ref{sec:results_limits}). Free nuisance parameters are profiled. A further test lets the sources emit any mixture of $\nu_e$, $\nu_\mu$, and $\nu_\tau$. The mass splitting $\dmsq$ does not enter.}
  \centering
  \footnotesize
  \renewcommand{\arraystretch}{1.2}
  \setlength{\tabcolsep}{4pt}
  \begin{tabular}{lll}
    \toprule
    Parameter & Default analysis & Variants \\
    \midrule
    $\Utau$ & Of interest, in $[0, 1]$ & Same \\
    $\Ue$, $\Umu$ & Fixed at 0 & Free, with $\chi^2_{\rm terr}$ \\
    $\delta_{14}$, $\delta_{24}$ & Do not enter & Free \\
    $s_{12}^2$, $s_{23}^2$, $s_{13}^2$, $\delta_{\rm CP}$ & Free, with $\chi^2_{\rm std}$ & Same \\
    Mass ordering & Normal & Inverted \\
    $\fS{e}$ & Free, in $[0, 1 - \fS{\tau}]$ & Same \\
    $\fS{\mu}$ & $1 - \fS{e} - \fS{\tau}$ & Same \\
    $\fS{\tau}$ & Fixed at 0 & Free, $\le 0.05$ or $0.10$ \\
    $\fS{s}$ & Fixed at 0 & Same \\
    \bottomrule
  \end{tabular}
\end{table}

\smallskip
\textit{Pull terms.}---The term $\chi^2_{\rm std}$ constrains the standard mixing parameters to their measured values. For IceCube MESE~2025, it sums the two-dimensional projections of the NuFIT~6.1 $\chi^2$ in $(s_{13}^2, s_{12}^2)$ and $(s_{23}^2, \delta_{\rm CP})$, for normal mass ordering~\cite{Esteban:2024eli, NuFIT61}; other pairings of the parameters yield the same limit. For 2040 and 2050, by which time oscillation experiments starting or under construction will have measured them more precisely, it is a sum of Gaussians whose centers are the NuFIT~6.1 best-fit values and whose widths are those projected in \Ref~\cite{Song:2020nfh}:
\begin{equation}
  \label{equ:chi2std}
  \chi^2_{\rm std} \!=\!
  \begin{cases}
    \Delta\chi^2_{\rm NuFIT}(s_{13}^2, s_{12}^2) & \\
    \quad {} + \Delta\chi^2_{\rm NuFIT}(s_{23}^2, \delta_{\rm CP}) & \text{(today)} , \\[3pt]
    \sum_k \left[ (\vartheta_k - \bar\vartheta_k)/\sigma_k \right]^2 & \text{(2040, 2050)} ,
  \end{cases}
\end{equation}
where $\bar\vartheta_k$ and $\sigma_k$ are the central value and the projected uncertainty of the parameter $\vartheta_k$, with $\sigma_k = 0.0016$, 0.0042, and 0.00059 for $s_{12}^2$, $s_{23}^2$, and $s_{13}^2$, and $6.7^\circ$ for $\delta_{\rm CP}$, for both 2040 and 2050.

The term $\chi^2_{\rm terr}$ enters only in the variant in which all three active-sterile mixings vary (Sec.~\ref{sec:results_limits}). It imposes terrestrial limits for averaged oscillations, $\Ue < 0.017$ and $\Umu < 0.024$ at 90\%~C.L., derived from the bounds on non-unitary mixing of \Ref~\cite{Blennow:2025qgd}. Each limit enters as a penalty that grows quadratically with the mixing and equals $\Delta\chi^2 = 1.64$, the asymptotic one-sided 90\%~C.L. threshold (Appendix~\ref{app:statistics_levels}), at the terrestrial limit:
\begin{equation}
  \label{equ:chi2terr}
  \chi^2_{\rm terr} = 1.64 \left[ \left( \frac{\Ue}{0.017} \right)^2 + \left( \frac{\Umu}{0.024} \right)^2 \right] .
\end{equation}

\subsection{Limits and discovery}
\label{sec:statistics_limits}

\textit{Test statistics.}---To constrain one active-sterile mixing, we minimize $\chi^2$ over the nuisance parameters (\tabl{parameters}) for each value of that mixing. Subtracting the global minimum, $\chi^2_{\min}$, gives the profile of that mixing; for $\Utau$, it is $\Delta\chi^2(\Utau)$. For upper limits, we use the one-sided test statistic of \Ref~\cite{Cowan:2010js}, which is zero for values of $\Utau$ below its best-fit value, $\widehat{\Utau}$:
\begin{equation}
  \label{equ:q_limit}
  q(\Utau) =
  \begin{cases}
    \Delta\chi^2(\Utau) & \Utau \ge \widehat{\Utau} , \\
    0 & \text{otherwise} .
  \end{cases}
\end{equation}
The upper limit is the value of $\Utau$ above which $q$ exceeds 2.71, the asymptotic 90\%~C.L.~threshold for one parameter~\cite{Wilks:1938dza}. We use it to compare our limits with the terrestrial limits of \figu{money_plot}, also at 90\%~C.L. Because $q$ is one-sided, this threshold is conservative (Appendix~\ref{app:statistics_levels}).

To test whether a measurement rejects standard mixing, we use the discovery test statistic~\cite{Cowan:2010js}
\begin{equation}
  \label{equ:q0}
  q_0 = \Delta\chi^2(\Utau = 0) , \quad
  Z = \Phi^{-1}(1 - p) \simeq \sqrt{q_0} ,
\end{equation}
where $p$ is the probability that standard mixing yields a $q_0$ at least as large as the one found, and $\Phi$ is the cumulative standard normal distribution. Asymptotically, $q_0$ is zero half of the time and follows a $\chi^2$ distribution with one degree of freedom otherwise~\cite{Wilks:1938dza, Chernoff:1954eli, Cowan:2010js}, so that $Z = \sqrt{q_0}$. For $\Utau$, we compute $p$ with pseudo-experiments instead, because the projected likelihoods are not Gaussian and, for muon-damped sources, $\fS{e}$ sits at the edge of its range. For $\Ue$ and $\Umu$, we use the asymptotic relation, since terrestrial limits at the eV scale already disfavor the values that flavor measurements could discover (Appendix~\ref{app:statistics_reach}).

Section~\ref{sec:measurements_projections} centers each projected likelihood on the flavor composition that its source yields with standard mixing. To derive limits, we use these likelihoods as they are, \ie, we assume that the future measurement finds exactly that composition. To compute the discovery significance for a given value of $\Utau$, we instead center the likelihood on the composition that this value yields. Limits and significances obtained in this way---from the Asimov data set~\cite{Cowan:2010js}---are close to their medians over many repeated measurements (Appendices~\ref{app:statistics_toys} and \ref{app:statistics_reach}).

\smallskip
\textit{Pseudo-experiments.}---We check the coverage of our limits and calibrate the discovery significance with pseudo-experiments (Appendix~\ref{app:statistics_translation}). Each simulates a flavor measurement, by IceCube MESE~2025 or a projected detector combination. We shift the composition at Earth that a chosen true value of $\Utau$ yields by a random offset, whose spread matches the uncertainty of the measurement. Since we know each likelihood only around one composition (\figu{likelihoods}), we translate it onto the shifted composition, assuming that a measurement there would yield a likelihood of the same shape. (We tested this assumption by simulating the forecasts at three other compositions. For the main 2050 combination, the 68\% and 90\%~C.L. intervals on $\Utau$ from these likelihoods agree within 5\% with those from translated ones.) Fitting the translated likelihood gives an upper limit on $\Utau$. The limit exceeds the true value in at least 94\% of the pseudo-experiments for IceCube MESE~2025 and in at least 96\% for the projections, more often than the 90\% that we quote. Thus, our limits are conservative (Appendix~\ref{app:statistics_toys}).

\section{Results}
\label{sec:results}

\tabl{limits} and \figu{limit_bars} show our upper limits on $\Utau$, from IceCube MESE~2025 and from the projected flavor measurements of each detector combination of \tabl{detectors}.

\subsection{Current limits}
\label{sec:results_current}

The best fit of IceCube MESE~2025, $\Utau = 0.09$, is preferred over standard mixing by only $\Delta\chi^2 = 0.03$, so the present data agree with standard mixing. They disfavor $\Utau > 0.88$ at 90\%~C.L. (\figu{limit_bars}). The limit is weak because MESE allows a wide range of the $\nu_\tau$ fraction, $\fE{\tau} = 0.33^{+0.24}_{-0.25}$ (Sec.~\ref{sec:measurements_present}).
Only sterile mixing that removes most of the $\nu_\tau$ is disfavored: at $\Utau = 0.88$, $\fE{\tau}$ falls from 0.33 to 0.06 for pion-decay sources.

The limit depends on how the measurement is modeled and on what the sources emit. With a single power law for the energy spectrum---not the broken one we use by default---it tightens to 0.65 (Sec.~\ref{sec:measurements_present}). With up to 5\% of $\nu_\tau$ at the sources, it disappears: no value of $\Utau$ below 1 is disfavored. Letting $\Ue$ and $\Umu$ vary within their terrestrial limits, or assuming inverted mass ordering, changes it by at most 1\%. 

Today's precision is also too low to discover sterile mixing: even if $\Utau$ were 0.95, a measurement like IceCube MESE~2025 would reject standard mixing at only $1.8\sigma$ (computed with $Z = \sqrt{q_0}$; Sec.~\ref{sec:statistics_limits}).

\subsection{Projected limits}
\label{sec:results_limits}

\begin{figure}[t]
  \centering
  \includegraphics[width=\columnwidth]{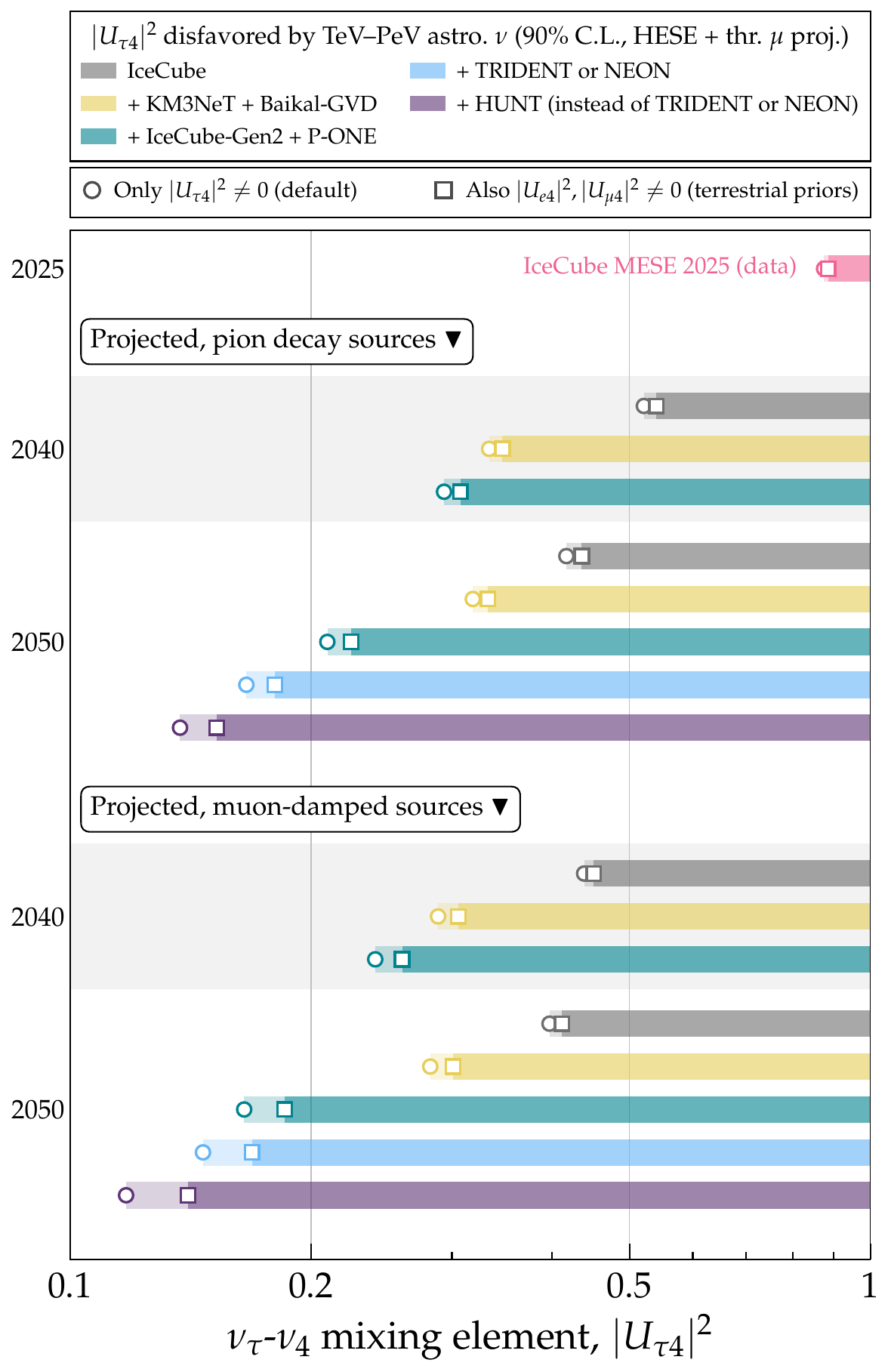}
  \caption{\label{fig:limit_bars}\textbf{\textit{Upper limits on $\Utau$ by detector combination.}} At 90\%~C.L. ($\Delta\chi^2 = 2.71$), from IceCube MESE~2025 and from the projected flavor measurements of the detector combinations of \tabl{detectors}, for pion-decay and muon-damped sources. Each bar spans the disfavored range: its dark part is disfavored also when $\Ue$ and $\Umu$ vary within their terrestrial limits, its light part only in the default analysis. \tabl{limits} lists the values. From IceCube alone in 2040 to the main combination in 2050, the limit tightens by nearly a factor of four for both source flavor compositions (Sec.~\ref{sec:results_limits}).}
  \vspace*{-0.2cm}
\end{figure}

\tabl{limits} and \figu{limit_bars} show that the projected limits tighten as more detectors are combined. For 2040, most of the gain comes from combining IceCube with KM3NeT and Baikal-GVD. From 2040 to 2050, the limits of the main combinations halve. About half of this gain comes from ten more years of KM3NeT, Baikal-GVD, P-ONE, and IceCube-Gen2 data, the rest from HUNT. The limits for muon-damped sources are 5--21\% tighter, in part because their larger standard $\nu_\tau$ fraction, 0.38 vs.~0.33, makes the same sterile mixing remove more $\nu_\tau$ [\equ{ftau}].

\begin{table*}[t]
  \caption{\label{tab:limits}\textbf{\textit{Upper limits on $\Utau$ at 90\%~C.L.}} At $\Delta\chi^2 = 2.71$, from IceCube MESE~2025 and the projections of \tabl{detectors}; within each year, each row adds detectors to the one above it. The projected limits are Asimov limits; the ranges in parentheses hold the central 68\% of the limits from pseudo-experiments (Sec.~\ref{sec:statistics_limits}). In the default analysis, only $\Utau$ is nonzero and the sources emit no $\nu_\tau$. In its variants (\tabl{parameters}), $\Ue$ and $\Umu$ also vary within their terrestrial limits (all mixings) or the sources emit up to 10\% of $\nu_\tau$ ($\fS{\tau} \le 0.10$). None: no value of $\Utau$ is disfavored. Bold: the main combinations. Each variant weakens the projected limits by at most a third.}
  \centering
  \footnotesize
  \renewcommand{\arraystretch}{1.15}
  \setlength{\tabcolsep}{6pt}
  \begin{tabular}{llcccccc}
    \toprule
    & & \multicolumn{3}{c}{Pion-decay sources} & \multicolumn{3}{c}{Muon-damped sources} \\
    \cmidrule(lr){3-5} \cmidrule(lr){6-8}
    Year & Detectors & Default & All mixings & $\fS{\tau} \le 0.10$ & Default & All mixings & $\fS{\tau} \le 0.10$ \\
    \midrule
    2025 & IceCube MESE~2025 (data) & 0.88 & 0.89 & None & 0.88 & 0.89 & None \\
    \midrule
    2040 & IceCube & 0.52 & 0.54 & 0.69 & 0.44 & 0.45 & 0.52 \\
     & + KM3NeT + Baikal-GVD & 0.33 (0.19--0.53) & 0.35 & 0.39 & 0.29 (0.19--0.44) & 0.31 & 0.32 \\
     & + IceCube-Gen2 + P-ONE & \textbf{0.29} (0.19--0.49) & \textbf{0.31} & \textbf{0.33} & \textbf{0.24} (0.18--0.40) & \textbf{0.26} & \textbf{0.27} \\
    \midrule
    2050 & IceCube & 0.42 & 0.44 & 0.51 & 0.40 & 0.41 & 0.46 \\
     & + KM3NeT + Baikal-GVD & 0.32 & 0.33 & 0.37 & 0.28 & 0.30 & 0.31 \\
     & + IceCube-Gen2 + P-ONE & 0.21 & 0.22 & 0.23 & 0.17 & 0.19 & 0.18 \\
     & + TRIDENT or NEON & 0.17 (0.10--0.26) & 0.18 & 0.18 & 0.15 (0.10--0.24) & 0.17 & 0.16 \\
     & + HUNT\footnote{Instead of TRIDENT or NEON.} & \textbf{0.14} (0.08--0.23) & \textbf{0.15} & \textbf{0.15} & \textbf{0.12} (0.07--0.18) & \textbf{0.14} & \textbf{0.13} \\
    \bottomrule
  \end{tabular}
\end{table*}

Analysis variants weaken the projected limits only moderately (\tabl{limits}). Letting $\Ue$ and $\Umu$ vary within their terrestrial limits [\equ{chi2terr}] weakens them by 3--20\% (\figu{limit_bars}). Letting the sources emit up to 5\% of $\nu_\tau$ weakens them by 4--11\%; up to 10\%, by 9--32\%, because these $\nu_\tau$ partly offset the deficit (Sec.~\ref{sec:flavor_tau}). Inverted mass ordering changes them by less than 1\%. Our limits do rely on the sources emitting few $\nu_\tau$. If the sources could emit any mixture of $\nu_e$, $\nu_\mu$, and $\nu_\tau$, the $\nu_\tau$ that they emit could offset any deficit. Flavor measurements would then set no limit.

Knowing the $\nu_e$ fraction at the sources to within $\pm 0.1$ would leave the main 2050 limits unchanged. Knowing it to within $\pm 0.01$ would tighten them to 0.08--0.10; exact knowledge would tighten them to 0.065--0.079. Such precision would require identifying the sources and how they make neutrinos. With the sources unknown, reaching the DeepCore limit, $\Utau < 0.054$~\cite{IceCube:2024dlz, IceCube:2025dcdata}, would need a measurement 2.3--2.6 times more precise than projected.

\subsection{Comparison with terrestrial limits}
\label{sec:results_terrestrial}

Figure~\ref{fig:money_plot} compares our limits with terrestrial limits. By 2040, our projected limits would be stronger than the MINOS limit, but weaker than the other terrestrial limits over almost all of their quoted ranges of $\dmsq$. By 2050, they would be stronger than the Super-K and MINOS limits and similar to the limit of the global fit of \Ref~\cite{Dentler:2018sju}. Our limits hold at all values of $\dmsq$ in the figure, since the $\nu_\tau$ deficit is independent of the mass of $\nu_4$, apart from matter effects inside the Earth (Sec.~\ref{sec:flavor_tau}).

The range of each terrestrial limit is set by the oscillation lengths or matter effects on which it relies. We draw the DeepCore limit~\cite{IceCube:2024dlz} at $\Umu = 0$, from its data release~\cite{IceCube:2025dcdata}, and the global fit of \Ref~\cite{Dentler:2018sju} at the values of $\dmsq$ at which it is reported. MINOS quotes its limit for $\nu_4$ nearly degenerate with $\nu_1$~\cite{MINOS:2011ysd}, which we take as $\dmsq < 10^{-4}$~eV$^2$, too small to drive oscillations over its baseline. We draw it for $\theta_{13} = 11.5^\circ$, since its stronger limit assumes $\theta_{13} = 0$, which current data exclude. The limits of MINOS, NO$\nu$A, DUNE, and ORCA on $\sin^2\theta_{34}$ also bound $\Utau$, because $\Utau = c_{14}^2 c_{24}^2 s_{34}^2 \leq \sin^2\theta_{34}$ (Sec.~\ref{sec:flavor_mixing}). Several terrestrial limits remain stronger than our 2050 limits: the DeepCore and ORCA6~\cite{KM3NeT:2025iar} limits at the eV scale, the NO$\nu$A limit~\cite{NOvA:2026ksk} between $3 \times 10^{-3}$ and 0.5~eV$^2$, and the projected DUNE limit~\cite{DUNE:2020fgq}. Cosmology disfavors $\Utau \gtrsim 10^{-3}$ between $10^{-2}$ and $10^2$~eV$^2$, but only for a standard cosmological thermal history~\cite{Hagstotz:2020ukm}.

Two factors keep astrophysical flavor measurements behind the strongest terrestrial limits. First, the $\nu_\tau$ deficit is small compared with the measurement precision. Sterile mixing at the level of today's terrestrial limits, $\Utau = 0.05$--0.2, lowers $\fE{\tau}$ by only 0.01--0.05 [\equ{ftau}], whereas, in the 68\%~C.L. regions of the main projections, $\fE{\tau}$ spans $\pm 0.07$ by 2040 and $\pm 0.03$ by 2050. Second, the source composition is unknown, so a change of its $\nu_e$ fraction can partly mimic the deficit. Terrestrial experiments avoid both factors: they detect many more neutrinos, whose flavor at production is known.

\subsection{Discovery prospects}
\label{sec:results_discovery}

The main 2050 combination could discover sterile mixing with $\nu_\tau$, \ie, reject standard mixing at $3\sigma$ in the median, if $\Utau \geq 0.23$ for pion-decay sources or $\Utau \geq 0.14$ for muon-damped ones (\figu{discovery_reach}). The main 2040 combination would reach $3\sigma$ only for muon-damped sources, if $\Utau \geq 0.27$. Terrestrial experiments already disfavor mixing this large for $\dmsq \gtrsim 3 \times 10^{-3}$~eV$^2$ (\figu{money_plot}), so a discovery would point either to a smaller $\dmsq$, where the quoted terrestrial limits are weaker or absent, or to a tension with terrestrial experiments. If $\Utau = 0.15$, the main 2050 combination would measure it with a relative uncertainty of 35--50\% at 68\%~C.L., larger for pion-decay sources.

\subsection{Other active-sterile mixings and the $\nu_\tau$ excess}
\label{sec:results_other}

Flavor measurements constrain the other two active-sterile mixings much less than $\Utau$ (Appendix~\ref{app:mixings}). They set no limit on $\Ue$, because the free source composition absorbs its effect. For $\Umu$, the main projections disfavor values above 0.37--0.62 and, for muon-damped sources, also intermediate values (Appendix~\ref{app:mixings}). Terrestrial limits~\cite{Blennow:2025qgd} already exclude all of these values. IceCube MESE~2025 also shows no $\nu_\tau$ excess, which would signal $\nu_\tau$ or $\nu_s$ emitted at the sources, or new physics beyond averaged 3+1 mixing~\cite{Arguelles:2015dca, Rasmussen:2017ert}.

\section{Summary and outlook}
\label{sec:summary}

Searches for a sterile neutrino, prompted by anomalies in short-baseline experiments, have focused on its mixing with $\nu_e$ and $\nu_\mu$. Its mixing with $\nu_\tau$, $\Utau$, the $\nu_\tau$ content of the fourth mass eigenstate, $\nu_4$, is the least tested. We have shown that the flavor composition of high-energy astrophysical neutrinos tests it. Although the sources emit almost no $\nu_\tau$, about one third of the flux at Earth is $\nu_\tau$, because of oscillations. Sterile mixing would lower this $\nu_\tau$ flux by a fraction $\Utau$. Since the sources do not produce $\nu_4$, this $\nu_\tau$ deficit does not depend on the mass splitting $\dmsq$, except through Earth-matter effects.

We have inferred $\Utau$ from the flavor composition measured by IceCube today and from projections for 2040 and 2050 that combine IceCube, KM3NeT, and Baikal-GVD with planned telescopes, including IceCube-Gen2 and HUNT. In all cases, we left the $\nu_e$ fraction at the sources free and let the standard mixing parameters vary within their uncertainties. 

\textbf{Today, IceCube data allow values of $\Utau$ up to 0.88 at 90\%~C.L. Adding the data of planned telescopes, if they are built, could bring this limit down to 0.24--0.29 by 2040 and to 0.12--0.14 by 2050.} Nonzero $\Ue$ and $\Umu$ within their terrestrial limits, up to 10\% of $\nu_\tau$ at the sources, or inverted mass ordering weakens our limits by at most a third (Sec.~\ref{sec:results_limits}).

By 2050, these limits would surpass those of Super-K and MINOS and reach that of the global fit of terrestrial data, but not the strongest terrestrial limits (Sec.~\ref{sec:results_terrestrial}). A discovery would need $\Utau \geq 0.14$--0.23 by 2050, values that terrestrial experiments already disfavor above $\dmsq \approx 3 \times 10^{-3}$~eV$^2$, so it would point to a smaller $\dmsq$ or to a tension with them.

Two kinds of progress would extend the reach of flavor measurements. The first is a better separation of $\nu_e$ from $\nu_\tau$ in neutrino telescopes, since the $\nu_e$--$\nu_\tau$ degeneracy limits the precision of the $\nu_\tau$ fraction most (Sec.~\ref{sec:measurements_flavor}). The second is better knowledge of the sources (Sec.~\ref{sec:results_limits}).

Next-generation neutrino telescopes aim mainly to find and understand the sources of high-energy astrophysical neutrinos. Their flavor measurements would also test, at no extra cost, whether a sterile neutrino mixes with $\nu_\tau$. They would do so with neutrinos, energies, and baselines unlike those of terrestrial experiments, and so with systematic uncertainties different from theirs.

\begin{acknowledgments}
MB is supported by {\sc Villum Fonden} under project no.~29388. This work used the Tycho supercomputer hosted at the SCIENCE High Performance Computing Center at the University of Copenhagen. QL is supported by the Canada First Research Excellence Fund and the Natural Sciences and Engineering Research Council of Canada through the Arthur B. McDonald Canadian Astroparticle Physics Research Institute. During the preparation of this work, the authors used Claude for research assistance in code development, numerical auditing, figure preparation, and editorial revision. All scientific judgments, validation decisions, and responsibility for the manuscript remain with the authors.
\end{acknowledgments}

\appendix
\counterwithin{figure}{section}
\counterwithin{table}{section}
\renewcommand{\thefigure}{\thesection.\arabic{figure}}
\renewcommand{\thetable}{\thesection.\arabic{table}}
\renewcommand{\theHfigure}{app.\thesection.\arabic{figure}}
\renewcommand{\theHtable}{app.\thesection.\arabic{table}}

\section{Projected flavor likelihoods}
\label{app:likelihoods}

\tabl{detectors} lists the detector combinations of the projected measurements and the years of data from each detector. For each combination and for pion-decay and muon-damped sources, the forecasts of \Refs~\cite{Liu:2023flr, Bustamante:2026aur, Bustamante:2026zst, Barenboim:2026mvn} provide posterior samples of the flavor composition at Earth and of the normalization and spectral index of the astrophysical neutrino flux. They parametrize the flavor composition with two angles, $\phi$ and $\psi$, and sample $\sin^4\phi$ and $\cos 2\psi$ from uniform priors:
\begin{equation}
  \label{equ:coords}
  \begin{aligned}
    & \fE{e} = \sin^2\phi \, \cos^2\psi , \\
    & \fE{\mu} = \sin^2\phi \, \sin^2\psi , \\
    & \fE{\tau} = \cos^2\phi , \\
    & \left| \frac{\partial(\sin^4\phi, \cos 2\psi)}{\partial(\fE{e}, \fE{\tau})} \right| = 4 ,
  \end{aligned}
\end{equation}
where the last line is the Jacobian of the map from $(\sin^4\phi, \cos 2\psi)$ to $(\fE{e}, \fE{\tau})$. Because the Jacobian is constant, the prior is uniform over the flavor triangle. Hence, the posterior density of the flavor composition, marginalized over the normalization and spectral index, is proportional to the likelihood marginalized over them.

\begin{table}[t]
  \caption{\label{tab:detectors}\textbf{\textit{Years of data from each detector in each flavor measurement.}} The present measurement, IceCube MESE~2025, and the projected detector combinations for 2040 and 2050, which combine high-energy starting events and through-going muons. Detector sizes, relative to IceCube (IC): KM3NeT (KM3), Baikal-GVD (GVD), and P-ONE (P-O), $1 \times$~IC; IceCube-Gen2 (Gen2) and TRIDENT, $7.5 \times$~IC; NEON, $10 \times$~IC; HUNT, $30 \times$~IC~\cite{Schumacher:2025qca}. T/N: TRIDENT or NEON. Start, Stop: years in which each detector starts and stops taking data, following the timeline of \Refs~\cite{Schumacher:2025qca, Barenboim:2026mvn}. IceCube stops when IceCube-Gen2 starts; without IceCube-Gen2, it takes data until 2040 or 2050. The other detectors take data at least until 2050. Bold: the main combinations.}
  \centering
  \renewcommand{\arraystretch}{1.15}
  \setlength{\tabcolsep}{4.5pt}
  \small
  \begin{tabular}{lccccccc}
    \toprule
    Year & IC & KM3 & GVD & P-O & Gen2 & T/N & HUNT \\
    \midrule
    Start & 2011 & 2027 & 2025 & 2031 & 2035 & 2040 & 2040 \\
    Stop & 2035 & & & & & & \\
    \midrule
    2025 & 11.4 & -- & -- & -- & -- & -- & -- \\
    \multicolumn{8}{l}{\quad IceCube MESE, data} \\
    \midrule
    2040 & 29 & -- & -- & -- & -- & -- & -- \\
         & 29 & 13 & 15 & -- & -- & -- & -- \\
         & \textbf{24} & \textbf{13} & \textbf{15} & \textbf{9} & \textbf{5} & -- & -- \\
    \midrule
    2050 & 39 & -- & -- & -- & -- & -- & -- \\
         & 39 & 23 & 25 & -- & -- & -- & -- \\
         & 24 & 23 & 25 & 19 & 15 & -- & -- \\
         & 24 & 23 & 25 & 19 & 15 & 10 & -- \\
         & \textbf{24} & \textbf{23} & \textbf{25} & \textbf{19} & \textbf{15} & -- & \textbf{10} \\
    \bottomrule
  \end{tabular}
\end{table}

We estimate this density, $\hat p(\boldsymbol{f}_\oplus)$, from the samples with a kernel density estimate. The test statistic of each projected measurement is then
\begin{equation}
  \label{equ:ts_proj}
  \mathrm{TS}(\boldsymbol{f}_\oplus) = -2 \ln \left[ \hat p(\boldsymbol{f}_\oplus) / \hat p_{\max} \right] ,
\end{equation}
where $\hat p_{\max}$ is the maximum of $\hat p$ over the flavor triangle. This TS plays the role of the TS of Sec.~\ref{sec:measurements_flavor}, with the normalization and spectral index marginalized over instead of profiled. The maximum of each likelihood lies close to, but not exactly on, the composition expected with NuFIT~6.1~\cite{Esteban:2024eli, NuFIT61}, partly because each forecast was simulated with an earlier fit of the standard mixing parameters. We translate each likelihood rigidly onto this composition, $(0.33, 0.34, 0.33)$ for pion-decay sources and $(0.22, 0.41, 0.38)$ for muon-damped sources. The translations are below 0.02 in each flavor fraction, except for IceCube alone in 2040 with pion-decay sources (0.035).

\begin{figure*}[t]
  \centering
  \includegraphics[width=\textwidth, height=0.94\textheight, keepaspectratio]{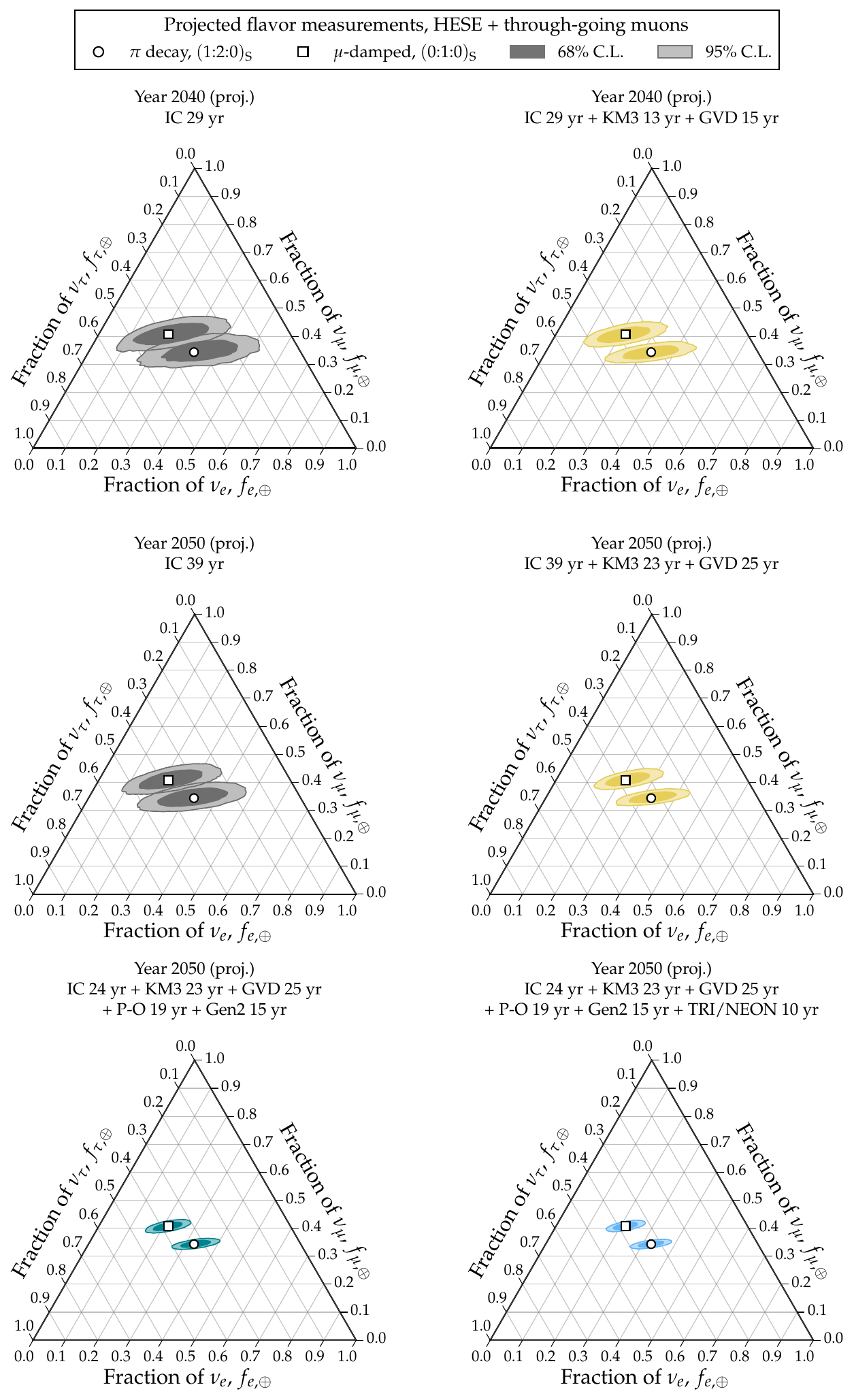}
  \caption{\label{fig:remaining_likelihoods}\textbf{\textit{Projected flavor-composition likelihoods of the other detector combinations.}} As \figu{likelihoods}, for the detector combinations of \tabl{detectors} not shown there. The 68\%~C.L.\ regions of the 2050 combination that includes TRIDENT or NEON are 9--11 times smaller in area than those of IceCube alone in 2040.}
\end{figure*}

Figure~\ref{fig:remaining_likelihoods} shows the likelihoods of the detector combinations of \tabl{detectors} that \figu{likelihoods} does not show. Their 68\%~C.L.\ regions shrink as more detectors are combined. Combining IceCube with KM3NeT and Baikal-GVD reduces the area of the region by a factor of 2.0--2.7. For 2050, further combining these three with IceCube-Gen2 and P-ONE reduces it by another factor of 2.3--2.4. In contrast, ten more years of IceCube alone, from 2040 to 2050, reduce it by a factor of only 1.3. The limits on $\Utau$ tighten most at the same two steps (\tabl{limits}).

\section{Statistical checks, discovery prospects}
\label{app:statistics}

\subsection{Limits at other confidence levels}
\label{app:statistics_levels}

Our limits in the main text, at $\Delta\chi^2 = 2.71$, are conservative because they are one-sided (Sec.~\ref{sec:statistics_limits}). Asymptotically, the interval of values of $\Utau$ with $\Delta\chi^2 < 2.71$ misses the true value in 10\% of repeated measurements~\cite{Wilks:1938dza}. In half of the misses, the best-fit value $\widehat{\Utau}$ lies too far above the true value; in the other half, too far below it. An upper limit misses the true value only in the second case, since it never lies below $\widehat{\Utau}$. Hence, asymptotically, our limits miss the true value in only 5\% of measurements. Limits that miss it in 10\% are set at $\Delta\chi^2 = 1.64$, the asymptotic one-sided 90\% threshold~\cite{Cowan:2010js}. Pseudo-experiments, which do not rely on asymptotic results, confirm that our limits cover the true value in more than 90\% of measurements (Appendix~\ref{app:statistics_toys}).

\tabl{limits95} lists the limits on $\Utau$ at $\Delta\chi^2 = 3.84$, the asymptotic threshold for a 95\%~C.L.~interval (97.5\% for one-sided limits), for the same detector combinations and analysis variants as \tabl{limits}. Limits at the asymptotic one-sided 90\% threshold, $\Delta\chi^2 = 1.64$, would be 19--24\% tighter than those that we quote for the main combinations and 12\% tighter for IceCube MESE~2025.

\begin{table*}[t]
  \caption{\label{tab:limits95}\textbf{\textit{Upper limits on $\Utau$ at 95\%~C.L.}} Same as \tabl{limits}, but at $\Delta\chi^2 = 3.84$ instead of 2.71. Raising the confidence level weakens the projected limits by 13--26\%.}
  \centering
  \footnotesize
  \renewcommand{\arraystretch}{1.15}
  \setlength{\tabcolsep}{6pt}
  \begin{tabular}{llcccccc}
    \toprule
    & & \multicolumn{3}{c}{Pion-decay sources} & \multicolumn{3}{c}{Muon-damped sources} \\
    \cmidrule(lr){3-5} \cmidrule(lr){6-8}
    Year & Detectors & Default & All mixings & $\fS{\tau} \le 0.10$ & Default & All mixings & $\fS{\tau} \le 0.10$ \\
    \midrule
    2025 & IceCube MESE~2025 (data) & 0.94 & 0.96 & None & 0.94 & 0.96 & None \\
    \midrule
    2040 & IceCube & 0.60 & 0.62 & 0.86 & 0.51 & 0.52 & 0.63 \\
     & + KM3NeT + Baikal-GVD & 0.40 (0.25--0.60) & 0.42 & 0.49 & 0.35 (0.24--0.50) & 0.36 & 0.40 \\
     & + IceCube-Gen2 + P-ONE & \textbf{0.33} (0.23--0.53) & \textbf{0.35} & \textbf{0.39} & \textbf{0.29} (0.22--0.45) & \textbf{0.31} & \textbf{0.33} \\
    \midrule
    2050 & IceCube & 0.48 & 0.50 & 0.61 & 0.45 & 0.46 & 0.54 \\
     & + KM3NeT + Baikal-GVD & 0.36 & 0.38 & 0.43 & 0.33 & 0.36 & 0.38 \\
     & + IceCube-Gen2 + P-ONE & 0.25 & 0.26 & 0.28 & 0.20 & 0.22 & 0.22 \\
     & + TRIDENT or NEON & 0.20 (0.12--0.29) & 0.21 & 0.22 & 0.17 (0.12--0.26) & 0.20 & 0.19 \\
     & + HUNT & \textbf{0.16} (0.11--0.25) & \textbf{0.18} & \textbf{0.18} & \textbf{0.14} (0.09--0.20) & \textbf{0.16} & \textbf{0.15} \\
    \bottomrule
  \end{tabular}
\end{table*}

\subsection{Pseudo-experiments}
\label{app:statistics_translation}

Each pseudo-experiment simulates one flavor measurement for a true value of $\Utau$, which yields the composition $\boldsymbol{f}_\oplus^{\rm true}$. We draw a composition, $\tilde{\boldsymbol{f}}$, from the likelihood of the measurement and add its offset from the peak, $\boldsymbol{f}_{\rm c}$, to $\boldsymbol{f}_\oplus^{\rm true}$. We translate the TS onto the resulting measured composition, $\boldsymbol{f}_\oplus^{\rm obs}$, which scatters around the true one as the likelihood does around its peak:
\begin{equation}
  \label{equ:toys}
  \begin{aligned}
    \boldsymbol{f}_{\oplus}^{\rm obs} &= \boldsymbol{f}_{\oplus}^{\rm true} + (\tilde{\boldsymbol{f}} - \boldsymbol{f}_{\rm c}) , \\
    \mathrm{TS}_{\rm toy}(\boldsymbol{f}_\oplus) &= \mathrm{TS}\!\left( \boldsymbol{f}_\oplus - \boldsymbol{f}_{\oplus}^{\rm obs} + \boldsymbol{f}_{\rm c} \right) .
  \end{aligned}
\end{equation}
For the projections, we draw $\tilde{\boldsymbol{f}}$ from the posterior samples of the forecast and the central values of the standard mixing parameters in $\chi^2_{\rm std}$ from their projected Gaussians. For IceCube MESE~2025, we draw $\tilde{\boldsymbol{f}}$ from the published TS grid. We fit each pseudo-experiment as we fit the measurements (Sec.~\ref{sec:statistics}).

Translating a projected likelihood assumes that its shape does not depend on the composition at which the measurement is simulated. To test this, we simulated the forecasts anew at the compositions that pion-decay sources yield with $\Utau = 0.42$, 0.49, and 0.56. These lie farther from the standard composition than those at which the main projections could discover sterile mixing at $3\sigma$: $\fE{\tau}$ is 0.11--0.15 below its standard value, instead of 0.03--0.07 (Sec.~\ref{sec:results_discovery}). For the main 2050 combination, the profiles of $\Utau$ from the simulated and the translated likelihoods have best fits within 0.01 of each other and 68\% and 90\%~C.L.~intervals within 5\%. For the main 2040 combination, the translation raises the significance with which standard mixing is rejected by 3--9\%.

\subsection{Coverage of the limits}
\label{app:statistics_toys}

Figure~\ref{fig:toy_checks}, right, shows the coverage of the upper limits of the main projections, \ie, the fraction of pseudo-experiments in which the limit lies above the true value of $\Utau$. We generated 1000 pseudo-experiments at each of seven true values, from 0.25 to 2 times the Asimov limit. At $\Delta\chi^2 = 2.71$ (our limits), the coverage is 97.6--100\%; at 1.64, it is 94.6--99.9\%, still above the asymptotic 90\%. For the other detector combinations and for the analysis variants (\tabl{parameters}), we generated 500 pseudo-experiments at each of three true values, 0.5, 1, and 1.5 times their limits. Their coverage at $\Delta\chi^2 = 2.71$ is at least 96\%.

The coverage exceeds the asymptotic one because the test statistic $q$ tends to be smaller than its asymptotic distribution predicts, for two reasons. First, when $q$ is nonzero, it is about half as large as a $\chi^2$ variable with one degree of freedom: for the main projections, its median is 0.22--0.36, instead of 0.45. Second, for muon-damped sources, the best fit of $\Utau$ exceeds the true value in 63--74\% of the pseudo-experiments, instead of in half of them, so $q$ is zero more often. Muon-damped sources emit no $\nu_e$, so their composition lies at one end of the segment of standard compositions. No change of $\fS{e}$ can absorb a fluctuation that moves the measured composition beyond that end, so the fit attributes part of it to sterile mixing. As a result, for the main projections, the value of $q$ that pseudo-experiments exceed with 10\% probability, which would give limits with exact 90\% coverage, is 0.4--1.0, below the asymptotic 1.64 and the 2.71 that we use.

For IceCube MESE~2025, we generated 1000 pseudo-experiments at each of six true values, from 0 to 0.9. As $\Utau$ grows, the distribution of $q$ approaches its asymptotic form. At 0.6--0.9, the coverage at $\Delta\chi^2 = 2.71$ is 94--95\%, close to the asymptotic 95\%; at 1.64, it is 89--90\%. Therefore, the limit from IceCube MESE~2025 is also conservative, although less so than the projected limits, whose coverage is at least 96\%.

For each projection, we also computed the upper limit on $\Utau$ in 300 pseudo-experiments without sterile mixing. The central 68\% of these limits gives the ranges in parentheses in Tables~\ref{tab:limits} and \ref{tab:limits95}. Their median lies within 16\% of the Asimov limit, so the limits that we quote are close to those of a typical measurement.

\subsection{Discovery significance}
\label{app:statistics_reach}

\begin{figure*}[t]
  \centering
  \includegraphics[width=\textwidth]{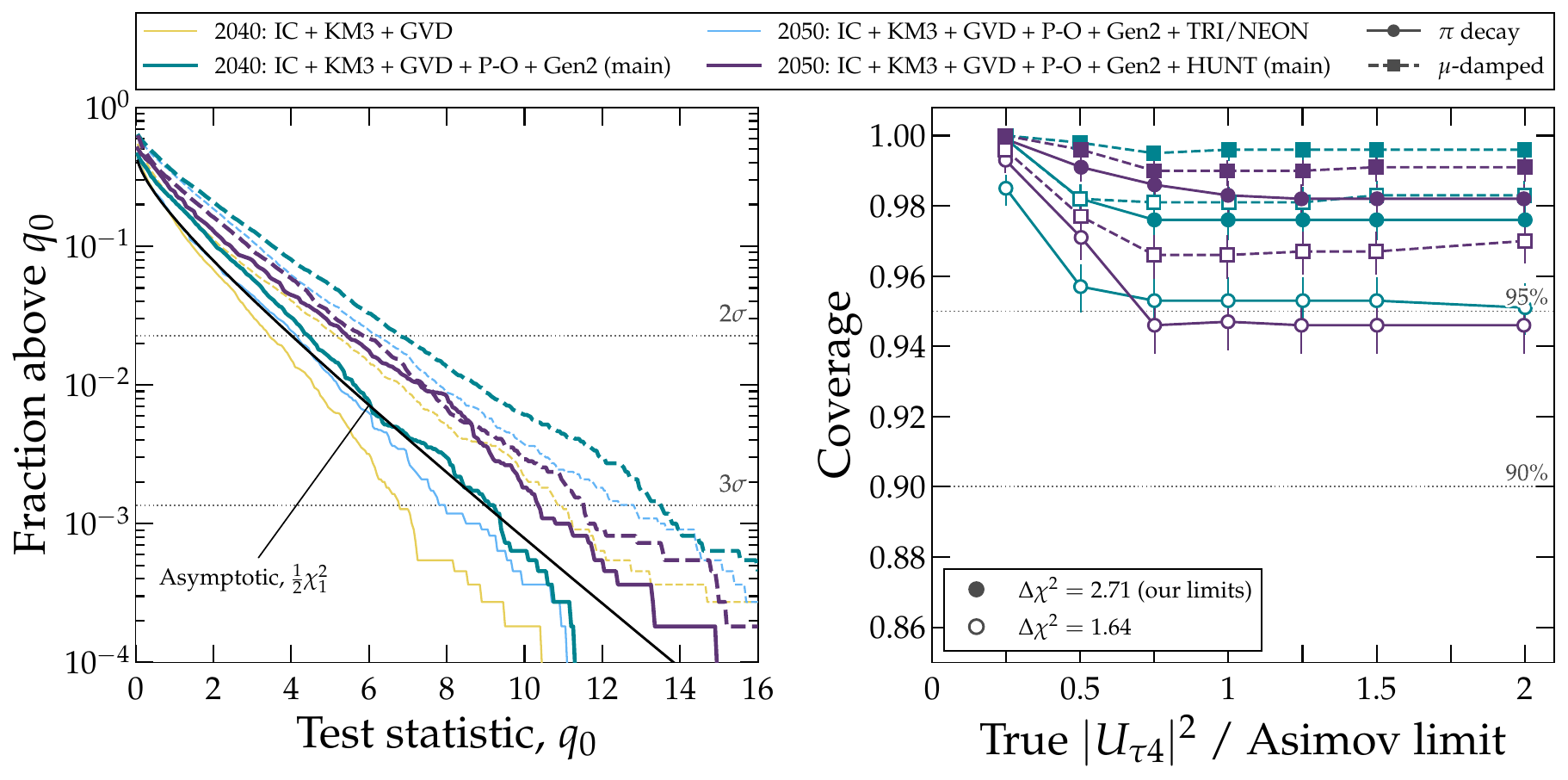}
  \caption{\label{fig:toy_checks}\textbf{\textit{Checks with pseudo-experiments.}} \textit{Left:} fraction of the 11000 pseudo-experiments without sterile mixing, per projection, in which $q_0$ exceeds a given value. Each curve crosses the $2\sigma$ and $3\sigma$ lines at the calibrated thresholds $q^*_2$ and $q^*_3$. \textit{Right:} coverage of the upper limits on $\Utau$ of the main projections, from 1000 pseudo-experiments per point; error bars are 68\% binomial uncertainties. The coverage of our limits exceeds the nominal 90\%~C.L.~(Appendix~\ref{app:statistics_toys}).}
\end{figure*}

Figure~\ref{fig:toy_checks}, left, shows the distribution of the discovery test statistic, $q_0$, in 11000 pseudo-experiments without sterile mixing for each projection. Its tail differs from the asymptotic one (Sec.~\ref{sec:statistics_limits}), so the asymptotic relation $Z = \sqrt{q_0}$ would overestimate the significance for some projections and underestimate it for others. Hence, we calibrate the significance with these pseudo-experiments. We define the threshold $q^*_n$ as the value of $q_0$ that these pseudo-experiments exceed with the probability of an $n\sigma$ fluctuation, $1 - \Phi(n)$, and convert $q_0$ into a significance by interpolating between these thresholds. Beyond $q^*_3$, which only about 15 pseudo-experiments exceed, we rescale the asymptotic relation to pass through $q^*_3$:
\begin{equation}
  \label{equ:z_calibrated}
  Z = 3 \sqrt{q_0 / q^*_3} \quad (q_0 > q^*_3) .
\end{equation}
For each projection, we also generated 500 pseudo-experiments at one value of $\Utau$, with an Asimov significance of 1.5--3.2$\sigma$. Their median significance lies within $0.25\sigma$ of the Asimov one, so the significances that we quote are close to those of a typical measurement.

\begin{figure*}[t]
  \centering
  \includegraphics[width=\textwidth]{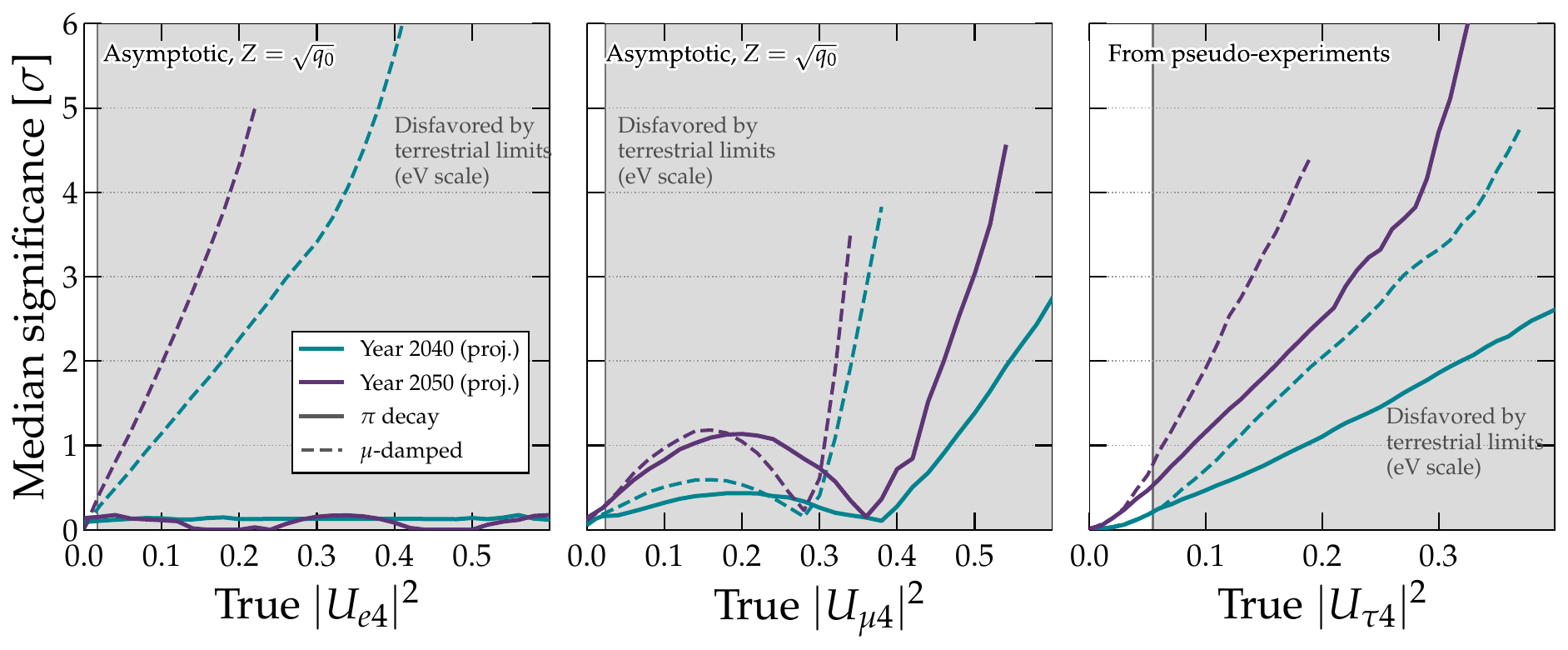}
  \caption{\label{fig:discovery_reach}\textbf{\textit{Discovery prospects.}} Median significance with which the main projected flavor measurements would reject standard mixing, against the true value of one active-sterile mixing, with the other two at zero. Curves end where the standard composition falls outside the region sampled by the forecasts. By 2050, sterile mixing with $\nu_\tau$ could reach $3\sigma$ for $\Utau \geq 0.14$--0.23, which terrestrial limits allow only for $\dmsq \lesssim 3 \times 10^{-3}$~eV$^2$ (Sec.~\ref{sec:results_discovery}).}
  \vspace*{-0.2cm}
\end{figure*}

Figure~\ref{fig:discovery_reach} shows the median significance with which the main projections would reject standard mixing, against the true value of each active-sterile mixing. Sterile mixing with $\nu_e$ moves the composition almost as a source with less $\nu_e$ would (Sec.~\ref{sec:flavor_see}). For pion-decay sources, a lower $\fS{e}$ absorbs this shift. For muon-damped sources, which emit no $\nu_e$, the shift moves the composition beyond the end of the segment of standard compositions, which no source reaches, so it is detectable. The significance from sterile mixing with $\nu_\mu$ peaks and then drops to nearly zero close to $\Umu_\star$, where the compensation described in Sec.~\ref{sec:flavor_see} is complete.

\section{Probabilities with one $|U_{\alpha 4}|^2$ mixing}
\label{app:single_angle}

Here we derive \equ{single_angle}, the averaged probabilities when only one active-sterile mixing angle, $\theta_{x4}$ with $x = e$, $\mu$, or $\tau$, is nonzero. The other two rotations $R_{y4}$ in \equ{mixing} are then the identity, so the mixing matrix reduces to $U = R_{x4} (U_3 \oplus 1)$, where $U_3 = R_{23} R_{13} R_{12}$ is the standard mixing matrix and $U_3 \oplus 1$ is the $4 \times 4$ matrix with $U_3$ as its upper-left block, 1 as its last diagonal entry, and zeros elsewhere. The rotation $R_{x4}$ mixes only the row of flavor $x$ with the sterile row. Since both the sterile row and the fourth column of $U_3 \oplus 1$ are $(0, 0, 0, 1)$, the active rows of the mixing matrix are
\begin{equation}
  \label{equ:single_angle_rows}
  \begin{aligned}
    U_{\alpha i} &= c_\alpha \, (U_3)_{\alpha i} \quad (i = 1, 2, 3) , \\
    U_{\alpha 4} &= \delta_{\alpha x} \, s_x \, e^{-i\delta_{x4}} ,
  \end{aligned}
\end{equation}
where $\alpha = e, \mu, \tau$, $s_x \equiv \sin\theta_{x4}$, $c_\alpha \equiv \cos\theta_{x4}$ for $\alpha = x$ and $c_\alpha \equiv 1$ otherwise, $\delta_{\alpha x}$ is the Kronecker delta, and $\delta_{x4}$ is the phase accompanying $\theta_{x4}$ (zero for $x = \tau$). Thus, $|U_{x4}|^2 = s_x^2$; in the parametrization of \equ{mixing}, this requires $\theta_{14} = 0$ for $x = \mu$ and $\theta_{14} = \theta_{24} = 0$ for $x = \tau$.

Substituting \equ{single_angle_rows} into \equ{prob} gives \equ{single_angle},
\begin{equation}
  \begin{aligned}
    P_{\alpha\beta} &= c_\alpha^2 c_\beta^2 \sum_{i=1}^{3} |(U_3)_{\alpha i}|^2 |(U_3)_{\beta i}|^2 + |U_{\alpha 4}|^2 |U_{\beta 4}|^2 \\
    &= c_\alpha^2 c_\beta^2 \, \Pstd_{\alpha\beta} + \delta_{\alpha x} \delta_{\beta x} \, s_x^4 ,
  \end{aligned}
\end{equation}
where $\Pstd_{\alpha\beta}$ is the probability with standard mixing.

\section{The \texorpdfstring{$\nu_\tau$}{nu-tau} deficit}
\label{app:tau_deficit}

Here we derive the $\nu_\tau$ deficit of Sec.~\ref{sec:flavor_tau}. Consider a source $(\fS{e}, 1-\fS{e}, 0)_{\rm S}$, which emits neither $\nu_\tau$ nor $\nu_s$. Sterile mixing with $\nu_\tau$ alone multiplies $P_{e\tau}$ and $P_{\mu\tau}$ by $1 - \Utau$ and leaves $P_{e e}$, $P_{e\mu}$, $P_{\mu e}$, and $P_{\mu\mu}$ unchanged [\equ{single_angle}]. Hence, of the fractions $F_\beta$ in \equ{fearth}, it multiplies $F_\tau$ by $1 - \Utau$ and leaves $F_e$ and $F_\mu$ unchanged. With standard mixing, $F_s = 0$, so $F_e + F_\mu + F_\tau = 1$ and each $F_\beta$ equals the standard flavor fraction $f^{\rm std}_{\beta,\oplus}$. Normalizing as in \equ{fearth} then gives
\begin{equation}
  \label{equ:ftau}
  \fE{\tau} = \frac{(1 - \Utau) f^{\rm std}_{\tau,\oplus}}{1 - \Utau f^{\rm std}_{\tau,\oplus}} ,
\end{equation}
while $\fE{e}/\fE{\mu}$ keeps its standard value. 

The deficit of \equ{ftau} also appears if $\nu_4$ is too heavy to be produced at all, in which case the $3 \times 3$ active block of the mixing matrix is not unitary~\cite{Antusch:2006vwa, Fernandez-Martinez:2016lgt, Blennow:2025qgd}. If this block is written as $(1 - \alpha) U'$, with $U'$ unitary and $\alpha$ a lower-triangular matrix of small parameters~\cite{Xing:2007zj, Xing:2011ur, Escrihuela:2015wra, Blennow:2025qgd}, a nonzero $\alpha_{\tau\tau}$ alone multiplies $F_\tau$ by $(1 - \alpha_{\tau\tau})^2 \simeq 1 - 2\alpha_{\tau\tau}$, as sterile mixing with $\Utau \simeq 2\alpha_{\tau\tau}$ does.

\section{Directional flavor-composition shifts}
\label{app:directions}

Here we derive \equ{directions}. Apart from lowering all $P_{x\beta}$, which a change of source absorbs, and from the term $s_x^4$, sterile mixing with $\nu_x$ multiplies each fraction $F_\beta$ by $w_\beta = 1 - |U_{x4}|^2 \delta_{\beta x}$ [\equ{single_angle}]. Dividing the numerator and the denominator of \equ{fearth} by $\sum_\gamma F_\gamma$ gives the flavor fractions with the weight, $\fE{\beta} + \delta\fE{\beta}$, in terms of those without it, $\fE{\beta} = F_\beta / \sum_\gamma F_\gamma$:
\begin{equation}
  \label{equ:weighted_fractions}
  \fE{\beta} + \delta\fE{\beta} = \frac{w_\beta F_\beta}{\sum_\gamma w_\gamma F_\gamma}
  = \frac{(1 - |U_{x4}|^2 \delta_{\beta x}) \, \fE{\beta}}{1 - |U_{x4}|^2 \fE{x}} ,
\end{equation}
whose expansion to first order in $|U_{x4}|^2$ is \equ{directions}.

At the pion-decay composition, with the NuFIT~6.1 best fit~\cite{Esteban:2024eli, NuFIT61}, the shift in \equ{directions} makes an angle with the segment of standard compositions (Sec.~\ref{sec:flavor_averaged}) of $4.6^\circ$ for $x = e$ and $55^\circ$ for $x = \mu$. These angles are larger than those of the curves in \figu{mixing_directions} (Sec.~\ref{sec:flavor_see}), because the curves are drawn for a fixed source: lowering all $P_{x\beta}$ then adds a shift along the segment.

\begin{figure*}[t!]
  \renewcommand{\thefigure}{\protect\ref*{app:mixings}.\arabic{figure}}
  \centering
  \includegraphics[width=\textwidth]{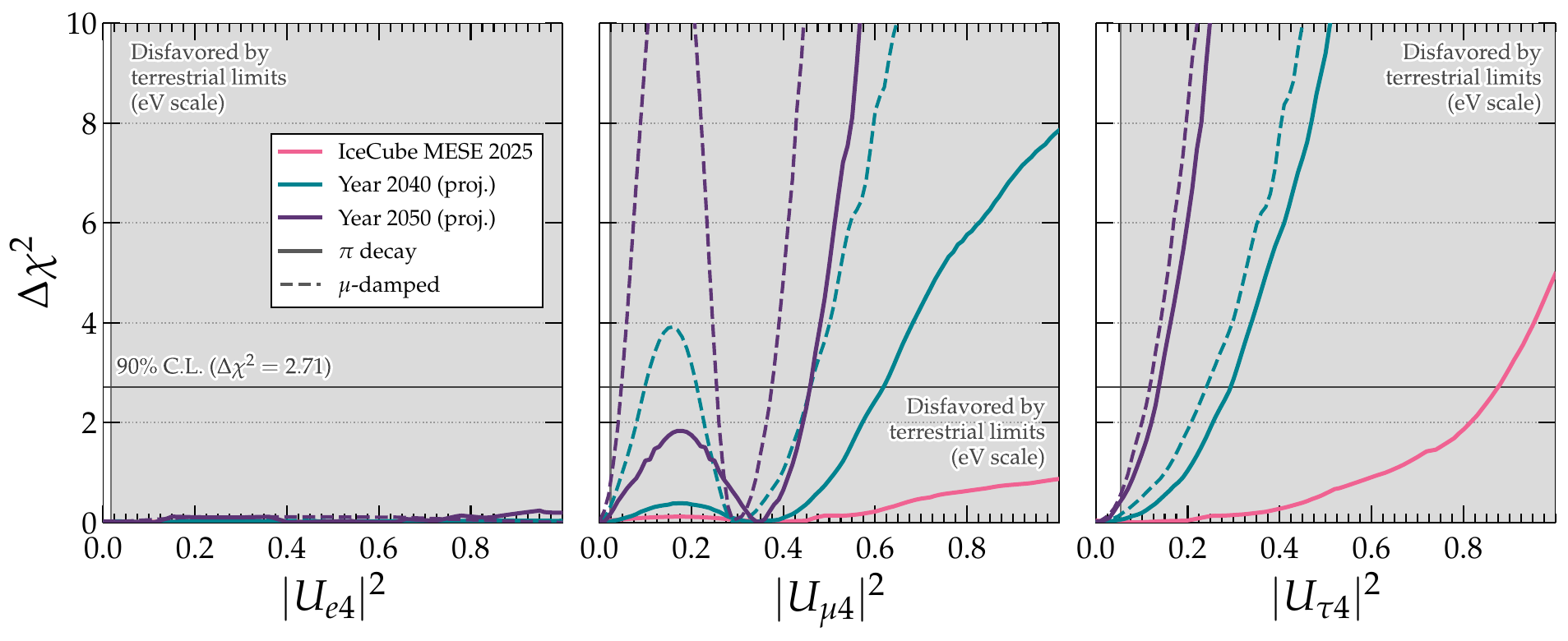}
  \caption{\label{fig:profiles}\textbf{\textit{Profiles of the three active-sterile mixings.}} For IceCube MESE~2025 and the main projections, with one active-sterile mixing nonzero at a time, profiled over the standard mixing parameters and the source composition. Only the profile of $\Utau$ grows steadily. The profile of $\Ue$ stays flat; that of $\Umu$ falls back to nearly zero close to $\Umu_\star$ (Sec.~\ref{sec:flavor_see}).}
  \vspace*{-0.2cm}
\end{figure*}

\section{Sterile mixing with \texorpdfstring{$\nu_e$}{nu-e} or \texorpdfstring{$\nu_\mu$}{nu-mu} only}
\label{app:emu}

When only $\nu_x$, with $x = e$ or $\mu$, mixes with $\nu_4$, \equ{single_angle} gives, for a source $(\fS{e}, \fS{\mu}, 0)$,
\begin{equation}
  \label{equ:emu_fractions}
  F_\beta = N c_\beta^2 \, F_\beta^{\rm std}(f'_{e,{\rm S}}) + \delta_{\beta x} \, \fS{x} \, |U_{x4}|^4 ,
\end{equation}
where $N \equiv \sum_\alpha \fS{\alpha} c_\alpha^2 = 1 - \fS{x} |U_{x4}|^2$, $f'_{e,{\rm S}} \equiv \fS{e} c_e^2 / N$, and $F_\beta^{\rm std}(f'_{e,{\rm S}}) = f'_{e,{\rm S}} \Pstd_{e\beta} + (1 - f'_{e,{\rm S}}) \Pstd_{\mu\beta}$ is the value of $F_\beta$ with standard mixing for a source $(f'_{e,{\rm S}}, 1 - f'_{e,{\rm S}}, 0)$. Since $N$ cancels in the flavor fractions, the mixing has three effects: a change of the $\nu_e$ fraction at the source from $\fS{e}$ to $f'_{e,{\rm S}}$, a factor $c_x^2$ on the probability of detecting a $\nu_x$, and a term of order $|U_{x4}|^4$ added to $F_x$. Our analysis absorbs the first effect, because it profiles $\fS{e}$ over $[0, 1]$, which $f'_{e,{\rm S}}$ also spans.

\subsection{Sterile mixing with \texorpdfstring{$\nu_e$}{nu-e}}
\label{app:ue}

For $x = e$, the term $|U_{e4}|^4$ in \equ{emu_fractions} is of second order in $\Ue$, so only the factor $c_e^2$ remains, which a change of $\fS{e}$ absorbs nearly fully (Sec.~\ref{sec:flavor_see}).

\subsection{Sterile mixing with \texorpdfstring{$\nu_\mu$}{nu-mu}}
\label{app:umu}

For $x = \mu$, the factor $c_\mu^2$ lowers $F_\mu$ by $N \Umu F_\mu^{\rm std}(f'_{e,{\rm S}})$, while the term $|U_{\mu4}|^4$ in \equ{emu_fractions} raises it by $(1 - \fS{e}) |U_{\mu4}|^4$, which grows faster with $\Umu$. The flavor composition equals the standard composition of a source with $\nu_e$ fraction $f^0_{e,{\rm S}}$ when two conditions hold. First, $f'_{e,{\rm S}} = f^0_{e,{\rm S}}$, which makes $F_e$ and $F_\tau$ proportional to their standard values. Second, the two changes in $F_\mu$ cancel:
\begin{equation}
  \label{equ:umu_cancel}
  (1 - \fS{e}) \, \Umu = N f^{\rm std}_{\mu,\oplus}(f^0_{e,{\rm S}}) ,
\end{equation}
where we have used $F_\mu^{\rm std}(f^0_{e,{\rm S}}) = f^{\rm std}_{\mu,\oplus}(f^0_{e,{\rm S}})$, since the fractions $F_\beta^{\rm std}$ sum to one. The first condition, $\fS{e} = f^0_{e,{\rm S}} N$, gives $N = (1 - \Umu)/(1 - f^0_{e,{\rm S}} \Umu)$. Substituting $N$ into \equ{umu_cancel} yields
\begin{equation}
  \label{equ:umu_star}
  \Umu_\star = \frac{f^{\rm std}_{\mu,\oplus}}{\fS{\mu} + f^{\rm std}_{\mu,\oplus}} ,
\end{equation}
where $\fS{\mu} = 1 - f^0_{e,{\rm S}}$ is the $\nu_\mu$ fraction of the source whose standard composition is reached. The source that reaches it, with $\fS{e} = f^0_{e,{\rm S}} (1 - \Umu_\star)/(1 - f^0_{e,{\rm S}} \Umu_\star)$, lies in $[0, f^0_{e,{\rm S}}]$. For pion-decay and muon-damped sources, $\fS{\mu} = 2/3$ and 1, and $f^{\rm std}_{\mu,\oplus} = 0.34$ and 0.41 (NuFIT~6.1 best fit~\cite{Esteban:2024eli, NuFIT61}), so $\Umu_\star = 0.34$ and 0.29.

\section{The three $|U_{\alpha 4}|^2$ mixings}
\label{app:mixings}

Figure~\ref{fig:profiles} shows the profiles of the three active-sterile mixings. The profile of $\Ue$ stays flat, because a change of $\fS{e}$ absorbs the shift that sterile mixing with $\nu_e$ causes (Sec.~\ref{sec:flavor_see}). The profile of $\Utau$ rises steadily: a change of $\fS{e}$ restores $\fE{\mu}$, but leaves a change in the split between $\fE{e}$ and $\fE{\tau}$ that grows with $\Utau$ (Sec.~\ref{sec:measurements_flavor}).

The profile of $\Umu$ depends on the source. For pion-decay sources, the projected flavor measurement likelihoods peak at $\fE{\mu} = 0.34$. Depending on $\Umu$, sterile mixing lowers or raises the $\fE{\mu}$ that these sources yield. Because they emit some $\nu_e$, the fit can bring $\fE{\mu}$ back to 0.34 at any $\Umu$ by choosing a source with less or more $\nu_e$. With this source, however, $\fE{e}$ and $\fE{\tau}$ differ from their measured values, by at most 0.03 below $\Umu_\star = 0.34$ [\equ{umu_star}], not at all at $\Umu_\star$, and by a growing amount above it. Since flavor measurements resolve the split between $\fE{e}$ and $\fE{\tau}$ poorly (Sec.~\ref{sec:measurements_flavor}), the profile stays low up to $\Umu_\star$ and rises above it.

For muon-damped sources, the projected flavor measurement likelihoods peak at $\fE{\mu} = 0.41$. If $\Umu$ is nonzero but smaller than $\Umu_\star = 0.29$, sterile mixing lowers the $\fE{\mu}$ that these sources yield, by up to 0.02 at $\Umu \approx 0.16$. Unlike for pion-decay sources, the fit cannot raise $\fE{\mu}$ back to 0.41 by choosing a source with less $\nu_e$, because muon-damped sources emit no $\nu_e$. Since $\fE{\mu}$ is the best-measured fraction, even this small deficit is detectable, so the profile rises and then falls back to zero at $\Umu_\star$, where the deficit vanishes. If $\Umu$ is larger than $\Umu_\star$, sterile mixing instead raises the $\fE{\mu}$ that these sources yield above 0.41. The fit can then bring $\fE{\mu}$ back down to 0.41 by choosing a source that emits some $\nu_e$. With this source, however, $\fE{e}$ is higher and $\fE{\tau}$ is lower than their measured values, by an amount that grows with $\Umu$, so the profile rises again.

\bibliography{refs}

\end{document}